\documentclass[twoside,twocolumn,english,superscriptaddres,showpacs,longbibliography,aps,prb]{revtex4-2}

\usepackage[T1]{fontenc}
\usepackage[utf8]{inputenc}
\usepackage[english]{babel}
\usepackage{bm}
\usepackage{amsmath}
\usepackage{amssymb}
\usepackage{graphicx}
\usepackage[unicode=true, pdfusetitle, bookmarks=true,bookmarksnumbered=true,bookmarksopen=true,bookmarksopenlevel=1,
breaklinks=false,pdfborder={0 0 0},pdfborderstyle={},backref=false,color links, citecolor=blue,linkcolor=red,urlcolor=blue] {hyperref}
\usepackage[nameinlink]{cleveref}
\Crefname{figure}{Fig.}{}

\makeatletter
\def\set@firstnote#1{%
 \@ifnum{\firstnote@num=#1\relax}{}{%
  \class@warn@end{Endnote numbers changed: rerun LaTeX}%
 }%
 \immediate\write\@mainaux{%
   \global\mathchardef\string\firstnote@num#1\relax
 }%
}%

\usepackage{braket}
\usepackage{bm}
\usepackage{tikz}
\usepackage{pgfplots}
\usepackage{pgf}
\usepackage{subfigure}
\usepackage{nicematrix}
\usepackage{diagbox}
\usepackage{orcidlink}

\renewcommand{\selectlanguage}[1]{}

\makeatother

\pgfplotsset{compat=1.18}
\begin{document}

\title{Homomorphism, substructure, and ideal: Elementary but rigorous aspects of renormalization group or hierarchical structure of topological orders}
\author{Yoshiki Fukusumi}
\affiliation{Physics Division, National Center for Theoretical Sciences, National Taiwan University, Taipei 106319, Taiwan}
\affiliation{Center for Theory and Computation, National Tsing Hua University, Hsinchu 300044, Taiwan}
\pacs{73.43.Lp, 71.10.Pm}
\author{Yuma Furuta}
\affiliation{Institute for Advanced Study, Kyushu University, Fukuoka, Japan}
\date{\today}
\begin{abstract}
We propose a general quantum Hamiltonian formalism of a renormalization group (RG) flow with an emphasis on generalized symmetry by interpreting the elementary relationship between homomorphism, quotient ring, and projection. In our formalism, the noninvertible nature of the ideal of a fusion ring realizing the generalized symmetry of an ultraviolet (UV) theory plays a fundamental role in determining condensation rules between anyons, resulting in the infrared (IR) theories. Our algebraic method applies to the domain wall problem in $2+1$ dimensional topologically ordered systems and the corresponding classification of $1+1$ dimensional gapped phase, for example. An ideal decomposition of a fusion ring provides a straightforward but strong constraint on the gapped phase with noninvertible symmetry and its symmetry-breaking (or emergent symmetry) patterns. Moreover, even in several specific homomorphisms connected under massless RG flows, less familiar homomorphisms appear, and we conjecture that they correspond to partially solvable models in recent literature. Our work demonstrates the fundamental significance of the abstract algebraic structure (beyond group structure), ideal, for the RG in physics. 

\end{abstract}

\maketitle

\section{introduction}
\label{introduction}

Classification of the topological orders (TOs) and their fusion rules (or fusion rings\cite{Fuchs:1993et}) between anyons is one of the most fundamental interests in contemporary condensed matter physics and related research fields\cite{Laughlin:1983fy,Wen:1989zg,Wen:1989iv,Moore:1991ks,Kitaev:2006lla}. Recently, the transformation law of the anyons in topologically ordered systems through a domain wall, which has been studied in the Kondo or impurity problem\cite{KONDO1970183,Affleck:1990by,Kane:1992xse,Kane:1992zza,Wong:1994np,Fendley_1995,Lesage:1998jky,Fendley_2006,Fendley_2007} traditionally, has captured attention in the fields as a potential classification framework of TOs more commonly (We note \cite{Affleck:1995ge,Saleur:1998hq,affleck2009quantumimpurityproblemscondensed} as useful references for traditional research directions. We also note older works in astrophysics\cite{Kibble:1976sj,Callan:1984sa} from a historical view). There exist mutually related ideas to study such settings, such as gapped or symmetry-preserving domain walls\cite{Kong:2013aya,Lan:2014uaa,kong2015boundarybulkrelationtopologicalorders,Wan:2016php,Kong:2017hcw,Kawahigashi:2015lxa,Kaidi:2021gbs}, Witt equivalence\cite{davydov2011structure,DavydovMugerNikshychOstrik+2013+135+177,Moller:2024xtt}, and renormalization group (RG) domain walls\cite{Brunner:2007ur,Gaiotto:2012np}. More practically, it is known that the theory containing a domain wall can explain the controversy of thermal Hall conductance in experimental settings\cite{Wang2017TopologicalOF,Mross_2018} and this is a consequence of the anomaly-inflow mechanism\cite{Callan:1984sa,Stone:2012ud,Son:2015xqa,Barkeshli:2015afa}. In other words, if anyons in one theory can smoothly transform into anyons in another theory, these two theories are connected in some sense, and a nontrivial domain wall between them can exist. In this sense, bulk edge correspondence\cite{Laughlin:1983fy,Moore:1991ks} or the correspondence between conformal field theory (CFT) and topological quantum field theory (TQFT)\cite{Witten:1988hf}, CFT/TQFT in short, should be relaxed to this connectivity or equivalence relations when applying the field-theoretic ideas to a realistic system as in experimental settings.

On the other hand, such connectivity between theories is a close connection to symmetry-protected topological order (SPT) \cite{haldane2016groundstatepropertiesantiferromagnetic,Haldane:1982rj,Haldane:1983ru,PhysRevB.34.6372,Schulz,Affleck:1986pq,Affleck:1987ch,Affleck:1987vf,Alcaraz_1989,Alcaraz_1989,MOshikawa_1992,Pollmann:2009ryx,Pollmann:2009mhk,Gu:2009dr,Chen:2010zpc,Chen:2011bcp} and its gapless analog called gapless SPT\cite{Scaffidi:2017ppg,Parker:2017wrd,Thorngren:2020wet,Li:2023knf} by the CFT/TQFT correspondence\cite{Witten:1988hf}. It has been gradually established that t'Hooft anomaly\cite{tHooft:1979rat} or Lieb-Shultz-Mattis anomaly\cite{Lieb:1961fr,Schultz:1964fv} classification governs such connectivity between theories\cite{Furuya:2015coa,Lecheminant:2015iga,Cho:2017fgz,Numasawa:2017crf,Kikuchi:2019ytf,Fukusumi_2022_c,Kikuchi:2022ipr,Fukusumi:2024ejk} and the corresponding phenomena has been studied numerically in \cite{Chepiga:2022ciu,Herviou:2023unm,Chepiga:2024hjd}. The most fundamental object in this classification is the integer spin simple current proposed by Schellekens and the collaborators\cite{Schellekens:1989uf,Schellekens:1990ys,Gato-Rivera:1990lxi,Gato-Rivera:1991bcq,Gato-Rivera:1991bqv,Kreuzer:1993tf,Fuchs:1996dd,Fuchs:1996rq} which initially appeared in the study of discrete torsion\cite{Vafa:1986wx,Vafa:1989ih}. If the product of two theories forms an integer (or half-integer) spin simple current analogous to a hadron, the product theory can be extended or gauged with respect to this nontrivial integer spin simple current. By applying the folding trick\cite{Wong:1994np}, the product or coupled theory generates a class of domain walls between the theories phenomenologically. Hence, one may expect a systematic construction of anyon transformation laws between theories in the same anomaly class under such anomaly classification by the integer spin simple current. However, the research has still been limited at this stage (Probably, the fundamental problem in this research direction is the appearance of mixing of chirality of anyons in the coupled theories. To our knowledge, this object has been studied in the context of the fractional quantum Hall effect in higher Landau levels or anti-Pfaffian states \cite {Levin_2007,Lee_2007,PhysRevLett.117.096802}. See also the discussions in \cite{Kong:2019cuu,Fukusumi:2024ejk}).

In this work, we demonstrate a general phenomenology realizing an RG as a projection. For the construction of this projection, a mathematical structure called \emph{ideal} plays a fundamental role. An ideal is fundamentally noninvertible and nonabelian (or nongroup-like). Moreover, it is also different from a subring, which commonly appears in physics literature. In principle, one can construct (infinitely) many new transformation laws of anyons by considering the ideal decomposition of existing fusion rings. More precisely, we study anyon transformation law or ring homomorphism between anyons in conformal field theories connected by massless RG flows\cite{Zamolodchikov:1987ti,Zamolodchikov:1987jf,Zamolodchikov:1989hfa}. Surprisingly, even in the simplest flow, such as the $SU(2)_{1} \times SU(2)_{1}$ Wess-Zumino-Witten (WZW) model to the $SU(2)_{2}$ WZW model and the flow from the tricritical Ising CFT to the Ising CFT, negative and fractional fusion coefficient, and unusual (or nonintegrable) flows appear inevitably. The unusual fusion coefficient has appeared recently in \cite{Zhao:2023wtg,Li:2023mmw,Li:2023knf,Fukusumi:2023psx,Fukusumi:2024cnl}, and the unusual flow seems to correspond to partially solvable systems in \cite{Matsui_2024,Matsui2024BoundaryDS,Katsura:2024lrn}. 

In the previous century, group theory and the corresponding symmetry have been studied widely in physics communities (See \cite{RevModPhys.51.591,Slansky:1981yr}, for example). The extension of this framework to the ring theory is known as generalized symmetry\cite{Cobanera:2012dc,Gaiotto:2014kfa}. Our approach establishes this research direction in the quantum Hamiltonian formalism by formulating the RGs with mathematically rigorous relations. We stress that the ideal and its generalizations play the most fundamental role in studying the condensations resulting in the appearance of emergent symmetries, and this is completely beyond group theory.

The rest of the manuscript is organized as follows. In section \ref{substructure}, we provide a general or mathematical understanding of massless RG flow and its implications for gapped phases in $1+1$ dimensions and for bulk states in topological orders in $2+1$ dimensions. This is the main section of this work, and we demonstrate the fundamental importance of further studies on the ideal structure in the fusion ring. The argument is purely algebraic, and one can generalize the argument to higher-dimensional systems. In section \ref{generating_ideal}, we introduce a way of constructing a series of ideals from a given ring corresponding to a CFT. The method itself is well known in mathematical literature, but its implications for theoretical physics are quite strong. This construction itself implies the existence of (infinitely) many new series of massless RGs from a given CFT. In section \ref{construction_homomorphism}, we provide several detailed constructions of homomorphisms between fusion rings appearing in CFTs connected under massless RG flows. We demonstrate the existence of nontrivial homomorphisms that are outside of existing integrable massless RG flow and propose its possible connection to partially solvable systems in recent literature\cite{Matsui_2024,Matsui2024BoundaryDS,Katsura:2024lrn}. Section. \ref{conclusion} is the concluding remarks of this manuscript, and we remark on a possible extension of our research and related phenomena that will be realizable in experimental settings. In Appendix \ref{ideal_gauge_fixing}, we present an alternative way to obtain a Hamiltonian realizing the massless RG by taking some limit. The resultant model can be interpreted as a gauge model involving the ideal structures. In Appendix \ref{coset_junction}, we provide a conjectural decomposition of CFTs related by massless RG, which will be useful in studying the homomorphism between models. 

\subsection{Summary of the main ideas}
In this work, we revisit existing formalisms with a less familiar (but traditional) way and formulate several frameworks in a rigorous algebraic way. Before going into details, we summarize the main claims in this manuscript.
\begin{itemize}
\item{A massless RG in the quantum Hamiltonian formalism can be formulated as a projection at the level of linear algebra. (Eq. \eqref{ideal_matrix} and Fig.\ref{homomorphism_coset}.)}
\item{The structure of ideals of a given fusion ring (or UV theory) produces a general method for generating massless RG and the resultant IR theories. (Fig. \ref{ideal_noninvertible})}
\item{We provide a rigorous interpretation of the sandwich construction in literature as implications of the massless RG to the massive RG. (Fig. \ref{massive_massless} and Eq. \eqref{algebraic_sandwich})}
\item{A rigorous definition of the unbroken and emergent symmetries and their dual relations has been shown (Eq.\eqref{definition_unbroken} and Eq. \eqref{relation_emergent-unbroken}).}
\item{Only by assuming the UV and IR fusion rings, the homomorphism is not unique (Section \ref{construction_homomorphism}).}
\end{itemize}
We stress that the construction of IR or emergent theory can be determined only by assuming the UV theory, and the systematic construction of such an IR theory has not been achieved in the literature. By applying the CFT/TQFT correspondence, the above points provide a classification of fusion rings of anyons, or SymTFTs in TOs. There exist several theoretical studies assuming the homomorphism (or functor) in the communities (See literature in \cite{Kong:2019cuu,Fukusumi:2024ejk}), but our work is the first general framework for constructing such homomorphisms.

\section{Substructure from homomorphism: Projection as renormalization group flow}
\label{substructure}
In this section, we denote general symmetry-based aspects of massless and massive RG flows. We mainly discuss $1+1$ dimensional CFTs or corresponding $2+1$ dimensional TOs as examples, but one can apply the same strategy to general models in physics in higher space-time dimensions. Our argument is elementary in contemporary mathematics, so we list a classic textbook as a reference\cite{atiyah1969introduction}. The general algebraic expression of the massless RG\cite{Zamolodchikov:1987ti,Zamolodchikov:1987jf,Zamolodchikov:1989hfa} in this work is the following ring homomorphism,
\begin{align}
\rho: \ \mathbf{A}_{(1)}&\rightarrow \mathbf{A}_{(2)}\\
\rho(a_{(1)}+b_{(1)})&=\rho(a_{(1)})+\rho(b_{(1)}), \\
\rho(a_{(1)}\times b_{(1)})&=\rho(a_{(1)})\times\rho(b_{(1)}), 
\end{align}
where $\mathbf{A}$ is an algebra describing generalized symmetries (or integral of motions) or underlying fusion category of anyons, the lower index $(1)$ and $(2)$ label the ultra-violet (UV) and infra-red (IR) theories, respectively, and $a, b$ are the algebraic objects. The above conditions represent that the structure of fusion is compatible or preserved under the application of domain walls, and the theory is transformed smoothly (or masslessly) into another theory. The corresponding argument has appeared evidently in \cite{Gaiotto:2012np} based on the earlier work \cite{Crnkovic:1989ug} and the folding trick \cite{Wong:1994np}. When interpreting this homomorphism as a defect between two theories, it is referred to as the RG domain wall, and it is expected to describe a massless RG\cite{Zamolodchikov:1987ti,Zamolodchikov:1987jf,Zamolodchikov:1989hfa}. We also note that the categorical structure corresponding to the ring homomorphism, called a functor, has been studied as gapped (or symmetry-preserving) domain walls in \cite{Lan:2014uaa,kong2015boundarybulkrelationtopologicalorders,Kong:2017hcw,Kaidi:2021gbs} with emphasis on the theoretical aspect of TQFTs or TOs. It should also be stressed that the underlying category theory should be generalized to $\mathbb{C}$-linear category, where $\mathbb{C}$ represents the complex number field, because the gapped domain wall can produce noninteger coefficients\cite{Zhao:2023wtg}. For readers interested in the underlying phenomenology of this phenomenon, we explain it in Appendix \ref{coset_junction} and note the related work by the first author\cite{Fukusumi:2025xrj}.

We provide an interpretation of the algebraic analog of a coset of $\mathbf{A}_{(1)}$ by $\text{Ker} \rho$ or the \emph{quotient ring} $\mathbf{A}_{(1)}/\text{Ker} \rho$ which can be constructed from the homomorphism $\rho$ corresponding to the massless RGs. We note that the idea of introducing a coset by group symmetry, known as orbifolding, has been studied widely\cite{Hamidi:1986vh,Dixon:1986qv,Dixon:1986jc,Vafa:1986wx,Dijkgraaf:1989hb,Tachikawa:2017gyf,Bhardwaj:2017xup} and is generalized to nonabelian symmetry\cite{Diatlyk:2023fwf,Lu:2025gpt}, but the quotient ring which we will discuss is different from them. The object, $\text{Ker}\rho$, dividing a ring is an \emph{ideal}, and a nontrivial ideal only forms a pseudo ring or rng without identity. (We cite \cite{berlyne2014idealtheoryringstranslation} as a recent English translation of the fundamental work by Noether \cite{Noether1921} to emphasize the fundamental importance of ideals in contemporary mathematics.) In other words, an ideal is intrinsically noninvertible or nonabelian (or non-group-like). From these properties, it should be distinguished from a subring, which is also familiar in physics literature. We also note that the symbol ``$/$" in this section is different from ``$/$" in coset CFTs\cite{Goddard:1984vk,Goddard:1984hg,Goddard:1986ee,Goddard:1988md}, whereas there exist some connections between the models as in \cite{LeClair:2001yp,Gaiotto:2012np} (We introduce a conjectural relationship between coset CFTs and massless RGs in the Appendix. \ref{coset_junction}). As far as we know, the term ideal appeared in physics literature in \cite{Gepner:1990gr,Fuchs:1993et,Douglas_2009,andersen2014fusionringsquantumgroups} as fusion ring ideal, in \cite{Bultinck:2015bot} in the study of matrix product operator, in \cite{Klos:2021gab} with a close connection to massless RG flows and in \cite{Schuster:2023bqe} in studying polynomial ring representations of fractonic systems\cite{Haah:2011drr,Vijay:2016phm}.   

We restrict our attention to abstract algebra, not to category theory, because the underlying category theories can be premodular fusion or superfusion categories, or even outside of them that are still under development. However, the basic fusion ring in this manuscript is well-defined and has been studied several times both in the mathematics and physics communities. In other words, our discussion can be a clue to constructing an unexplored fundamental class of categories. The fusion rings with negative or fractional coefficients are the corresponding structure, and they appeared in the study of fermionic string theories \cite{Ginsparg:1988ui} and more recently in \cite{Li:2023mmw,Li:2023knf}, for example. The corresponding fusion rings have been studied more generally by the first author in \cite{Fukusumi:2023psx,Fukusumi:2024cnl} with a close connection to the simple current extension by Schellekens and the collaborators\cite{Schellekens:1989uf,Schellekens:1990ys,Gato-Rivera:1990lxi,Gato-Rivera:1991bcq,Gato-Rivera:1991bqv,Kreuzer:1993tf,Fuchs:1996dd,Fuchs:1996rq}. 

We also note that the topological symmetry operator (or integral of motion) is an operator over $\mathbb{C}$\cite{Petkova:2000ip}. Hence, the quantum Hamiltonian formalism of the symmetry operator itself does not restrict the coefficients to nonnegative integer coefficients, which are only permitted in the fusion category. The appearance of a nonnegative integer coefficient is the consequence of the correspondence between the symmetry operator and the defects under the topological Wick rotation (or modular $S$ transformation in two-dimensional CFTs)\cite{Petkova:2000ip}\footnote{We thank Yuji Tachikawa, Jurgen Fuchs, and Shinichiro Yahagi for the related discussions.}. In other words, the fusion categories studied in (pure) mathematics have become physical research interests, by identifying the algebraic objects as defect or symmetry operator (or conserved charges) in the pioneering work \cite{Petkova:2000ip}. Moreover, by focusing on their realizations as symmetry operators, which form rings over the complex number field $\mathbb{C}$, the ideas have been \emph{generalized} to $\mathbb{C}$-linear categories. Readers who are familiar with particular types of fusion categories in physics may not be used to treating ring over $\mathbb{C}$, but this kind of structure, conserved charges in quantum Hamiltonian systems described by linear algebra, is more established in other sciences, including experimental and theoretical physics, and mathematics. Keeping in mind this fact, it is remarkable that the formalism in \cite{Petkova:2000ip} is based on the quantum Hamiltonian formalism. We also comment on the realizations as defects, which do not permit noninteger coefficients before algebraic objects, but they are not in the main scope of this paper. The distinctions between the defect and symmetry operators and the related problems have been studied in \cite{Belletete:2018eua,Belletete:2020gst,Seiberg:2023cdc,Seiberg:2024gek}.  

An ideal $\mathbf{I}_{(1)}$ of $\mathbf{A}_{(1)}$ is a subset (or linear subspace) of $\mathbf{A}_{(1)}$, satisfying the following property,
\begin{equation}
\begin{split}
&\text{If the relations  $a_{(1)}\in \mathbf{A}_{(1)}$ and  $s_{(1)}\in \mathbf{I}_{(1)}$ are satisfied,} \\ &\text{then the relation $a_{(1)}\times s_{(1)}\in \mathbf{I}_{(1)}$ holds. } 
\end{split}
\end{equation}
where $\times$ represents the fusion product (or just product in abstract algebra) of the ring $\mathbf{A}_{(1)}$.
If $\mathbf{I}_{(1)}$ includes identity object, the equality $\mathbf{I}_{(1)}=\mathbf{A}_{(1)}$ holds and this case is called trivial ideal. To obtain a nontrivial ideal, it is necessary to avoid the appearance of the identity object in the ideal. Hence, the objects in a nontrivial ideal become noninvertible by definition. In more mathematical terminology, the ring-like structure without identity is called a pseudo ring or rng.

\begin{figure}[htbp]
\begin{center}
\includegraphics[width=0.5\textwidth]{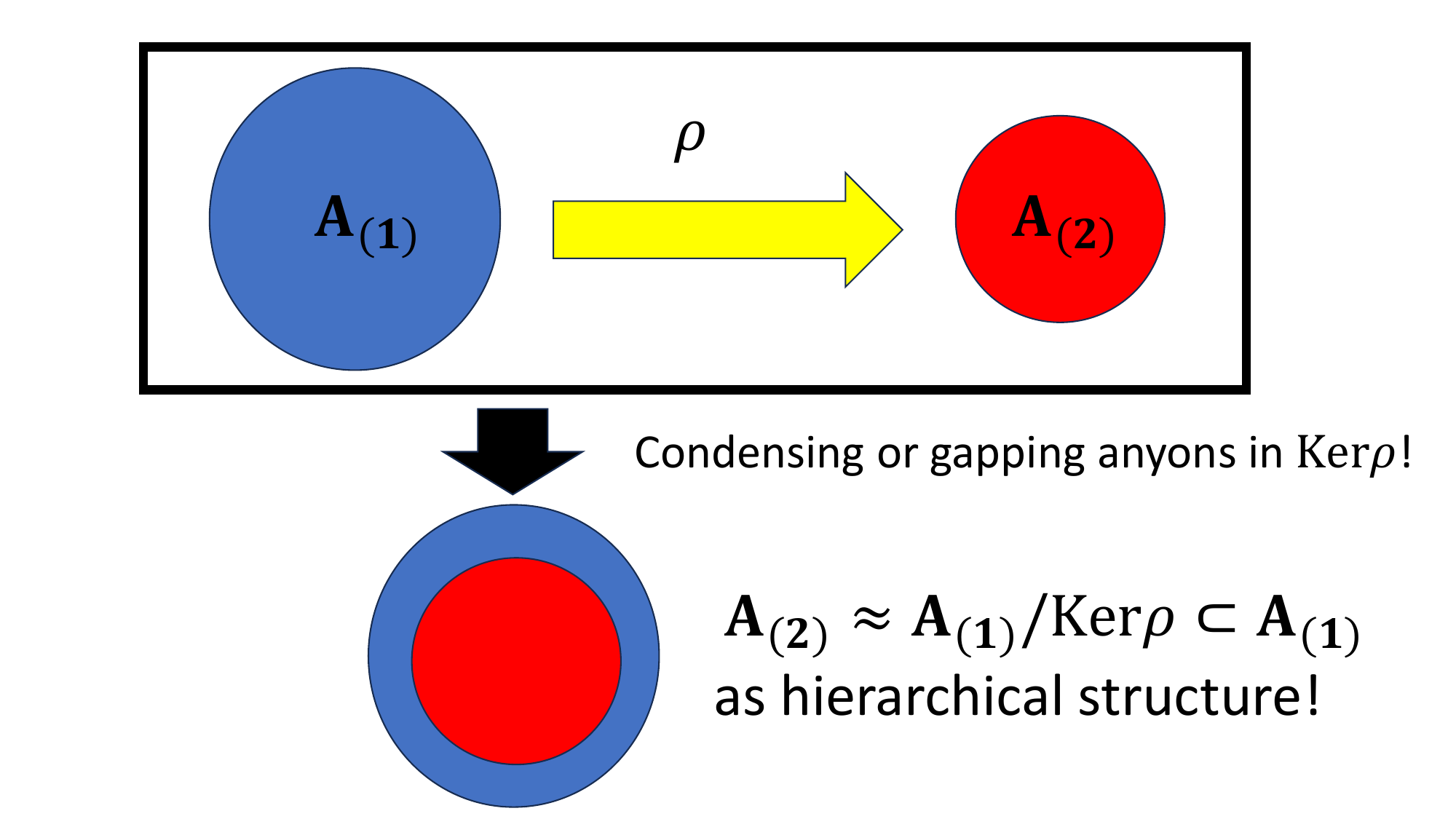}
\caption{Massless renormalization group and its connection to hierarchical structures in TOs. Under the CFT/TQFT, the homomorphism $\rho: \mathbf{A}_{(1)}\rightarrow \mathbf{A}_{(2)}$ induces hierarchical structure of anyons in TOs. Because of the existence of kernel of the homomorphism notified as $\text{Ker}\rho$ (which is intrinsically noninvertible), the fusion rule of a reduced theory $\mathbf{A}_{(2)}$ can become different from the original theory $\mathbf{A}_{(1)}$ and this can be understood as an emergent phenomenon\cite{Anderson:1972pca}(for more recent example, see the discussions and literature in \cite{Kikuchi:2021qxz,Kikuchi:2022gfi,Kikuchi:2022biw}). In other words, $\mathbf{A}_{(2)}$ can be different from a subring of $\mathbf{A}_{(1)}$. This figure can be seen as an algebraic representation of the hierarchical structure in fractional quantum Hall states\cite{Jain:1989tx,Bernevig2008PropertiesON,Bernevig_2008} and its connection to microscopic descriptions has been studied recently in \cite{Yang_2021,Yuzhu_2023}.}
\label{homomorphism_coset}
\end{center}
\end{figure}

When assuming the ring homomorphism is surjective, the following ring isomorphism (and inclusion) holds,
\begin{equation}
\mathbf{A}_{(2)}=\mathbf{A}_{(1)}/\text{Ker} \rho \ (\subset \mathbf{A}_{(1)})
\end{equation}
where $\text{Ker} \rho$, forming an ideal, is a kernel of the map $\rho$ and the inclusion is defined by interpreting the objects as linear algebra (not as a ring). This is widely known as the ring homomorphism theorem in mathematics (or an example of Noether's isomorphism theorems), and one can see the profound history initiated by Kummer, Dedekind, and more. In the present theoretical physics context, this means the correspondence between $\mathbf{A}_{(1)}/\text{Ker} \rho$ and the IR generalized symmetries (or integral of motions) of massless flow $\rho$. By treating both $\mathbf{A}_{(1)}$ and $\text{Ker} \rho$ as linear algebra, one can obtain the basis of $\mathbf{A}_{(1)}/\text{Ker} \rho$ straightfowardly. This corresponds to the hierarchical structure of a topologically ordered state by applying CFT/TQFT correspondence (FIG. \ref{homomorphism_coset}). The procedure taking the quotient is nothing but gapping anyons belonging to $\text{Ker} \rho$ or assuming ordering $s_{(1)}=0$ for all the elements $s_{(1)}\in\text{Ker} \rho$. Because of this operation, the algebraic relation is deformed to that of the IR theory, and the condition $s_{(1)}=0$ is the anyon or mass condensation. Moreover, by studying the massless RGs successively, one can obtain the sequence of subalgebra structures and the corresponding generalized symmetry-breaking pattern. In this discussion, the construction of the ring homomorphism from $\mathbf{A}_{(1)}$ is fundamental, and this can be constructed from the ideal decomposition of the ring $\mathbf{A}_{(1)}$. This argument can be thought of as a natural continuation of the research direction in \cite{Klos:2019axh,Klos:2020upw,Klos:2021gab}.

In the above discussion, we mainly considered its interpretation in the massless RG flows, but the same strategy can be applied to the classification of symmetry (or integral of motion) in the gapped phase.  First, let us consider the Hamiltonian $H_{(1)}$ with $\mathbf{A}_{(1)}$ symmetry. Then let us assume there exist a nontrivial ideal $\mathbf{I}_{(1)}$ where all the elements $\alpha_{(1)} \in \mathbf{I}_{(1)}$ are commutative. Then the corresponding topological operators satisfy the following conditions,
\begin{equation}
[H_{1}, Q_{s_{(1)}}]=0, \ [Q_{s'_{(1)}}, Q_{s_{(1)}}]=0,
\label{ideal_matrix}
\end{equation}
where $Q$ is a symmetry operator (or integral of motion) representing the symmetry action of $\mathbf{A}_{(1)}$, i.e. $Q_{s'_{(1)}}Q_{s_{(1)}}=Q_{s'_{(1)}\times s_{(1)}}$.
This means that the Hamiltonian and the symmetry operators (or integral of motions) are diagonalizable at the same time. Moreover, if $s_{(1)}$ is invertible, the ideal $\mathbf{I}_{(1)}$ becomes trivial because this means $\mathbf{A}_{(1)}=\mathbf{I}_{(1)}$. Hence, all the element $s_{(1)}\in \mathbf{I}_{(1)}$ is noninvertible. By interpreting the operator as a matrix\cite{Petkova:2000ip}, one can demonstrate that $Q_{s_{(1)}}$ has the eigenvalue $0$. Hence by projecting the theory with $Q_{s_{(1)}}=0$, one can obtain the Hamiltonian $H_{(2)}$ with $\mathbf{A}_{(1)}/\mathbf{I}_{(1)}=\mathbf{A}_{(2)}$ symmetry. We note several references to study such matrix representation of (topological) symmetry operators in lattice models \cite{Cobanera:2009as,Cobanera:2011wn,Cobanera:2012dc,Belletete:2018eua,Belletete:2020gst,Aasen:2020jwb,Lootens:2021tet,Lootens:2022avn,Lootens:2023wnl}. More practically, the careful treatment of emanant symmetry which distinguish the symmetry in lattice models and quantum field theoretic models (QFTs) will become important\cite{Cheng:2022sgb,Seiberg:2023cdc,Seiberg:2024gek}. For the later discussion, we stress that an abelian algebra $\mathbf{A}_{(1)}$ can have a nontrivial (or nonabelian) ideal because one treats $\mathbf{A}_{(1)}$ as a group ring in mathematical terminology. In other words, the notion of abelian or invertibility depends on the detailed settings and representations.

\begin{figure}[htbp]
\begin{center}
\includegraphics[width=0.5\textwidth]{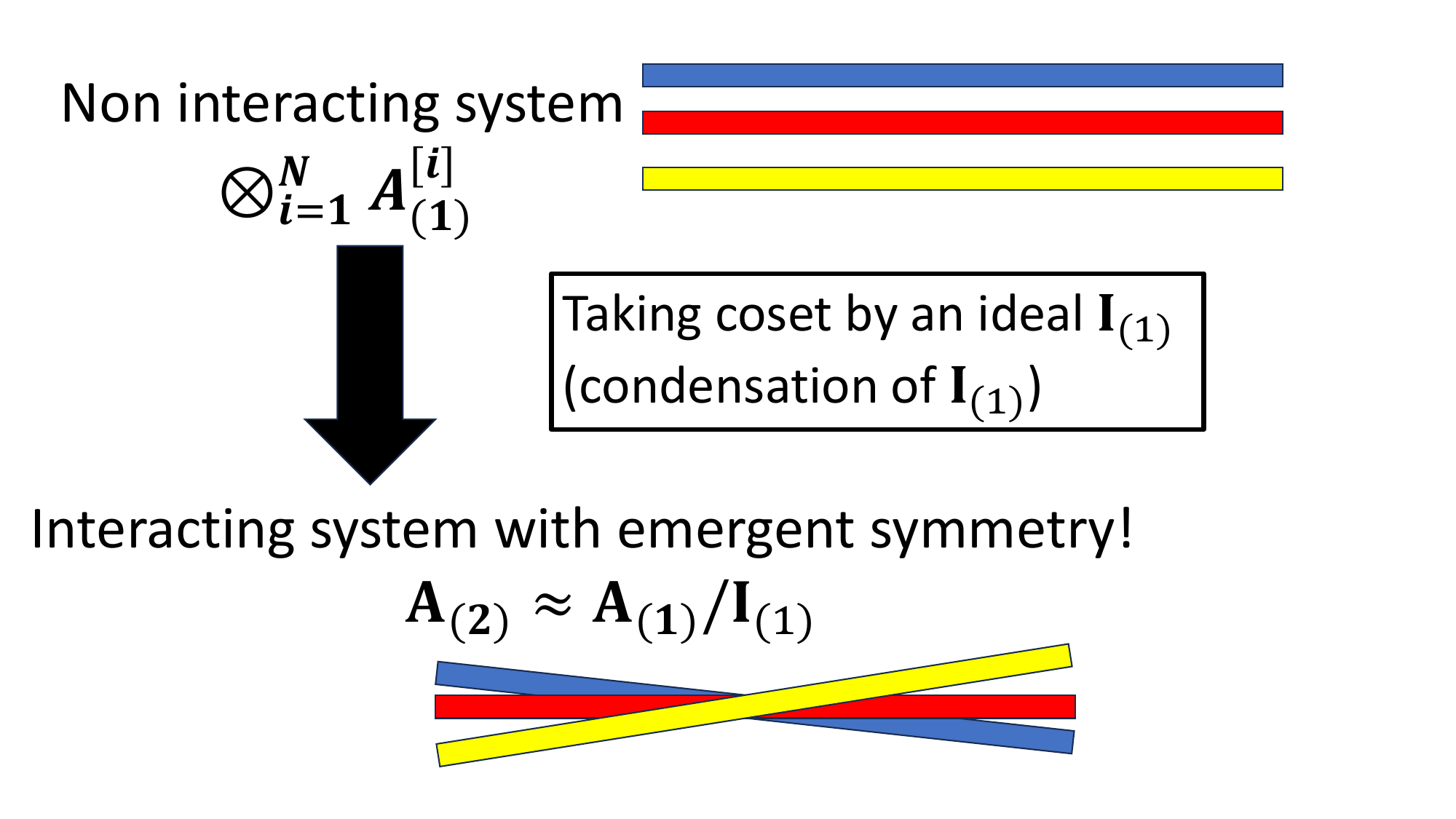}
\caption{Construction flow of emergent or enhanced symmetries from coupled symmetries. From the knowledge of existing theories $\mathbf{A}_{(1)}^{[i]}$ and the ideal of the coupled symmetry $\mathbf{A}_{(1)}$, one can obtain system with nontrivial or interacting symmetry $\mathbf{A}_{(2)}$. Only by assuming the existence of the commutativity of the kernel, one can realize the algebraic operation as a projection. Applications of this method to extended symmetry, such as Haagerup symmetry or magma in E-series models  \cite{Nivesvivat:2025odb} are interesting for future research. We note some research on the general construction of ideals in polynomial rings, which might be useful for the general construction of ideals\cite{GIANNI1988149,KAWAZOE20111158}. }
\label{algebraic_coupling}
\end{center}
\end{figure}

Based on the arguments above, we propose a systematic construction of systems with emergent symmetry (possibly noninvertible) from known existing lattice and QFT models (FIG. \ref{algebraic_coupling}). This will be applicable to coupled wire systems\cite{PhysRevLett.88.036401,Teo:2011hq,PhysRevB.91.245144,Klinovaja:2014iba,PhysRevB.91.085426,Kane:2017hld,Fuji_2017,Kane:2018amn,Goldman:2019wvz,Lim:2024qis} or modulated symmetric systems\cite{Chamon:2004lew,Haah:2011drr,Vijay:2016phm,Vijay:2016phm}, for example. The ideal decomposition of an abelian theory with a sufficient degree of freedom will be useful for the construction or classification, for example. Let us assume a Hamiltonian $H_{(1)}$ with the knowledge of symmetries $\mathbf{A}_{(1)}=\{ \mathbf{A}_{(1)}^{[i]}\}_{i=1}^{N}$ and its ideal $\mathbf{I}_{(1)}$. Then, one can obtain the Hamiltonian $H_{(2)}$ with symmetry $\mathbf{A}_{(1)}/\mathbf{I}_{(1)}$ by representing the quotient as a projection. For example, in a coupled wire model, the original model can be just a ladder model without inter-ladder interaction where the symmetry $\mathbf{A}_{(1)}^{[i]}$ acts on each ladder. However, by considering a nontrivial projection sending the ideal part to $0$, one can obtain a more nontrivial model with inter-ladder interaction, and the symmetry (or integral of motions) is transformed to $\mathbf{A}_{(2)}$. For a modulated symmetric system, an analogous argument based on gauging (or orbifolding) can be seen in \cite{Cao:2024qjj,Pace:2024tgk}, and corresponding arguments based on ideal will also be possible in principle.

 For the application of our method to the classification of the gapped phases, we remind that the massive RGs have a close connection to boundary conformal field theories (BCFTs)\cite{Date:1987zz,Saleur:1988zx,Qi_2012,Cardy:2017ufe,Foda:2017vog}. The correspondence has been studied in a wide variety of communities under various names as Moore-Seiberg data\cite{Moore:1988qv,Moore:1988ss,Moore:1989vd,Fuchs:2002cm}, Li-Haldane conjecture\cite{Li_2008,Qi_2012}, and boson condensation\cite{Bais:2008ni,Kong:2013aya}. The combination of our algebraic arguments and symmetry analysis of boundary state in \cite{Graham:2003nc} will provide the general criteria to describe the general class of gapped phases. Related arguments can be seen in \cite{Cho:2016xjw,Numasawa:2017crf,Kikuchi:2019ytf,Thorngren:2019iar,Li:2022drc,Cordova:2022lms,Kikuchi:2023gpj,Kikuchi:2023cgg}. However, as the author has clarified in \cite{Fukusumi:2024ejk}, the label of boundary state can be outside of the original symmetry $\mathbf{A}_{(1)}$ or $\mathbf{A}_{(2)}$. The corresponding boundary states are known as new boundary conditions in condensed matter\cite{Affleck:1998nq,Behrend:2000us,Iino:2020ipa} and symmetry-breaking boundary conditions in high energy or mathematical physics\cite{Birke:1999ik,Fuchs:1999xn,Quella:2002ct} because these boundary condition breaks extended bulk symmetry such as $W$ or Lie group symmetry in the bulk CFT (we note \cite{Recknagel:2013uja} as a useful reference). For the later convenience, we introduce the extended ring (or extended algebra) $\mathbf{A}^{\text{ex}}$ with $\mathbf{A}\subset\mathbf{A}^{\text{ex}}$ and assume that the underlying CFT Hamiltonians are symmetric under $\mathbf{A}^{\text{ex}}$ action but the Hilbert spaces are labeled by $\mathbf{A}^{\text{ex}}$. Then one can express the set $\mathbf{R}\subset \mathbf{A}^{\text{ex}}$ labelling $\mathbf{A'}$ invariant module as follows,
\begin{equation}
\begin{split}
&\text{If the relations  $a\in \mathbf{A'}$ and  $r \in \mathbf{R}(\subset \mathbf{A}^{\text{ex}})$ are satisfied,} \\ &\text{then the relation $a\times r\in \mathbf{R}$ holds. } 
\end{split}
\end{equation}
Hence, in this formalism, one can formulate $\mathbf{R}$ as a generalization of an ideal because $r \in \mathbf{R}$ can be outside of $\mathbf{A'}$. We propose that the above subring invariant set governs the massive RG and their classifications. We remind that whereas the set $\mathbf{R}$ is invariant under the action of the subring $\mathbf{A'}$, the object $r\in \mathbf{R}$ will be noninvariant, and this is nothing but the SSB.

\begin{figure}[htbp]
\begin{center}
\includegraphics[width=0.5\textwidth]{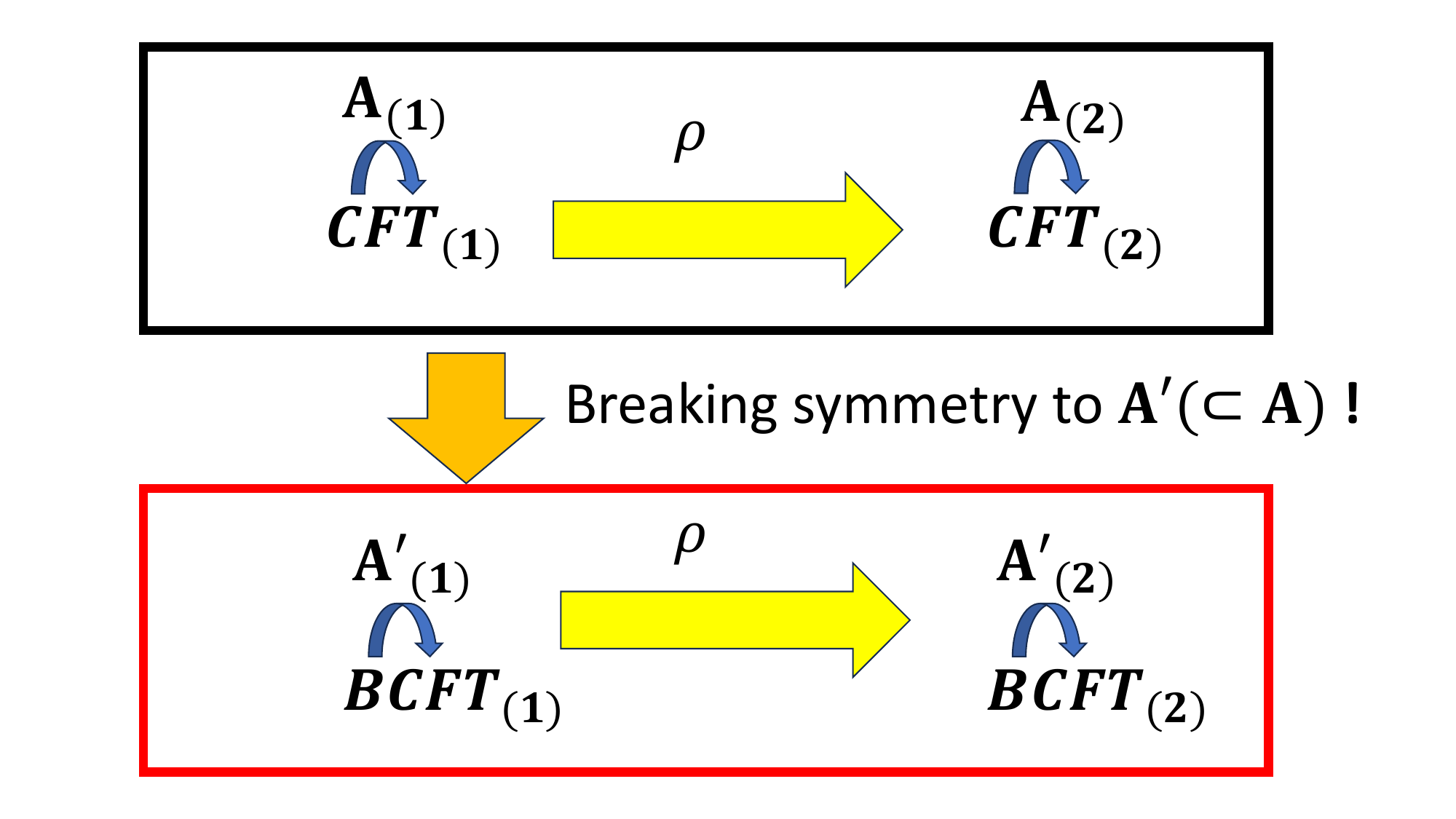}
\caption{Hierarchical structure of gapped phase induced from massless RG. The homomorphism $\rho$ induces natural hierarchical structures in $1+1$ dimensional gapped phase and corresponding $2+1$ dimensional bulk states of TOs. Assuming Moore-Seiberg data, one can label primary states by the object in $\mathbf{A}_{(1)}$ or $\mathbf{A}_{(2)}$. However, the label of the states in BCFT can be outside of $\mathbf{A}_{(1)}$ and called symmetry breaking boundary states or new boundary states and they will be labeled by extended algebra $\mathbf{A}_{(1)}^{\text{ex}}$ or $\mathbf{A}_{(2)}^{\text{ex}}$. Without such extensions, they will be labelled by objects in an ideal of $\mathbf{A'}_{(1)}$ or $\mathbf{A'}_{(2)}$.}
\label{massive_massless}
\end{center}
\end{figure}

When the system $(1)$ preserves the symmetry $\mathbf{A}'_{(1)} \subset \mathbf{A}_{(1)}^{\text{ex}}$ one can investigate possible Hilbert space as $\mathbf{A'}_{(1)}$ invariant Hilbert space or module (not states) spanned by these BCFTs. The spontaneous $\mathbf{A}'_{(1)}$ symmetry breaking corresponds to the appearance of $\mathbf{A}'_{(1)}$ noninvariant states. The existence of massless flow $\rho: \mathbf{A}_{(1)}\rightarrow \mathbf{A}_{(2)}$ implies the hierarchical relation of the symmetry of the gapped phase as $\rho: \mathbf{A}'_{(1)}\rightarrow \mathbf{A}'_{(2)}$ and that of the corresponding modules. By applying the argument of the construction of projection or embedding $\mathbf{A}_{(2)}\subset \mathbf{A}_{(1)}$, one can also obtain the corresponding projection for $\mathbf{A'}_{(2)}\subset \mathbf{A'}_{(1)}$ and corresponding hierarchy of gapped Hamiltonian. In other words, the massless RG almost automatically implies a hierarchical structure of $1+1$ dimensional gapped phases. Moreover, this provides hierarchical structures of entanglement spectrum or bulk states of $2+1$ dimensional topologically ordered systems by considering the CFT/TQFT correspondence\cite{Witten:1988hf,Moore:1988qv,Moore:1988ss,Moore:1989vd,Fuchs:2002cm,Li_2008,Bais:2008ni,Kong:2013aya} (FIG. \ref{massive_massless}).

If we assume the boundary states are labeled by the extended algebra $\mathbf{A}_{(1)}^{\text{ex}}$ with $\mathbf{A}_{(1)}\subset \mathbf{A}_{(1)}^{\text{ex}}$ and the scope of the homomorphism $\rho$ can be extended to the algebra  $\mathbf{A}_{(1)}^{\text{ex}}$ as in \cite{Affleck:1998nq,Behrend:2000us,Iino:2020ipa,Birke:1999ik,Fuchs:1999xn,Quella:2002ct}, one can define the action of $\rho$ or module homomorphism as follows,

\begin{equation} 
\rho(|\alpha_{(1)}\rangle)=|\rho(\alpha_{(1)})\rangle,
\end{equation}
where $\alpha_{(1)}$ is an element of the extended algebra $\mathbf{A}_{(1)}^{\text{ex}}$ and the states are corresponding boundary states. Hence by applying the method in \cite{Graham:2003nc}, one can obtain the action of $\mathbf{A}_{(2)}^{\text{ex}}=\text{Im}\rho_{|\mathbf{A}_{(1)}^{\text{ex}}}$ as follows,
\begin{equation} 
\rho(Q_{\beta_{(1)}})\rho(|\alpha_{(1)}\rangle)=|\rho(\beta_{(1)}\times\alpha_{(1)})\rangle
\end{equation}
where $\beta_{(1)}$ is an element of the ring $\mathbf{A}_{(1)}^{\text{ex}}$. By restricting $\beta_{(1)}\in \mathbf{A'}_{(1)}$ and choosing modules labeled by the elements of extended algebra $\mathbf{A}_{(1)}^{\text{ex}}$, one can obtain appropriate $\mathbf{A}'_{(2)}$ module labeled by the elements of the extended algebra $\mathbf{A}_{(2)}^{\text{ex}}$. This argument provides a symmetry-based formalism to characterize the hierarchical structure of the gapped phases, revealing the symmetry properties of Hilbert spaces. For further studies in this research direction, the construction of symmetry-breaking boundary conditions and the corresponding defects is fundamentally important. 

We also note that if one assumes that the BCFTs are labelled by $\mathbf{A'}$, one can observe that a $\mathbf{A'}$ module is labelled by objects in an ideal of $\mathbf{A'}$.
For the simplest example, one can obtain the trivial ideal $\mathbf{A'}$. In other words, 
\begin{equation}
\text{A gapped phase is an ideal of unbroken subring $\mathbf{A'}$.} 
\end{equation}
when assuming this modified version of the Moore-Seiberg data and massive RG flow. Inversely, $\mathbf{A'}$ invariant sets $\mathbf{R}$ in the extended ring $\mathbf{A}^{\text{ex}}$ can be thought of as a generalization of an ideal.
 
Based on the recent development of sandwich construction, one can interpret the set $\mathbf{R}$ as follows, in general,
\begin{equation}
\begin{split}
&\{\text{Label of $D$-dimensional gapped phase} \} \\
= &\{ \text{Condensable $D+1$-dimension symmetry}\} 
\end{split}
\label{algebraic_sandwich}
\end{equation}
where both $D$ and $D+1$ dimensional system has same symmetry described by a ring $\mathbf{A}'$.
The latter also determines the massless RG flows in $D$-dimensional systems with the UV symmetry $\mathbf{A}'$. In other words, the condensable objects in a $D+1$-dimensional system can appear in a $ D$-dimensional boundary as stable (or surviving) objects, and this phenomenon is governed by an ideal or its generalizations. This provides a more complete and general algebraic understanding of the sandwich construction proposed in \cite{Fukusumi:2024cnl}.

\subsection{Unbroken and emergent symmetry}

Based on the arguments in this section, we define unbroken (or exact) symmetry and emergent symmetry under an RG flow $\rho: \mathbf{A}_{(1)}\rightarrow \mathbf{A}_{(2)}$. The discussions in this subsection provide the unified definition distinguishing the unbroken and emergent symmetry with a complete and rigorous language. The emergent symmetry and the corresponding phenomena are the central interest of both condensed matter and high-energy physics\cite{Anderson:1972pca}, and our expression will apply to a large class of models when their symmetries are described by a ring.

A phenomenology of unbroken symmetry under RG flow is simple: a symmetry subring $\mathbf{A}_{(1)}^{\text{ub}}\subset \mathbf{A}_{(1)}$ of UV theory, which does not change under a RG flow $\rho$, is an unbroken symmetry. In the algebraic language, this condition means that $\mathbf{A}_{(1)}^{\text{ub}}$ and $\text{Ker} \rho$ are disjoint except for $0$. Hence, the following defines the unbroken symmetry,

\begin{equation}
\begin{split}
& \mathbf{A}_{(1)}^{\text{ub}} \text{ is an unbroken symmetry under } \rho   \\
&\leftrightarrow \mathbf{A}_{(1)}^{\text{ub}} \cap \text{Ker} \rho =\{0\}
\end{split}
\label{definition_unbroken}
\end{equation}

In other words, when the above condition is not satisfied, the symmetry $\mathbf{A}_{(1)}^{\text{ub}}$ is deformed or broken nontrivially under $\rho$. This general argument provides a testable and accessible criterion of emergent and exact symmetry and will be useful to test conjectural RG flow, for example, in \cite{Kikuchi:2021qxz,Kikuchi:2022gfi,Kikuchi:2022biw}. It should be remarked that as linear algebra, one can consider the basis of $\mathbf{A}_{(1)}/ \text{Ker} \rho$, but it does not imply that $\mathbf{A}_{(1)}/ \text{Ker}\rho$ forms subring of original ring $\mathbf{A}_{(1)}$. Hence, the dimension of unbroken symmetry (which should form a subalgebra) is less than the dimension of $\mathbf{A}_{(1)}/ \text{Ker} \rho$ and the following inequality holds,
\begin{equation}
\text{dim}\mathbf{A}_{(2)}\ge\text{dim}\mathbf{A}_{(1)}^{\text{ub}},
\end{equation}
where dim is the dimension of a ring in linear algebra. In other words, the dimension of the emergent symmetry $\mathbf{A}_{(2)}^{\text{em}}$ can be expressed as,
\begin{equation}
\text{dim}\mathbf{A}^{\text{em}}_{(2)}\le \text{dim}\mathbf{A}_{(2)}-\text{dim}\mathbf{A}_{(1)}^{\text{ub}},
\label{relation_emergent-unbroken}
\end{equation}

By taking a maximal unbroken subalgebra under $\rho$, one can restrict the form of perturbation in the UV theory, and the above inequality provides the constraint in comparing the IR (or emergent) theories. When taking $\mathbf{A}_{(1)}^{\text{ub}}$ a maximal unbroken subalgebra, the inequality will become equality in some cases. The resultant relation in these cases is,
\begin{equation}
\text{dim}\mathbf{A}^{\text{em}}_{(2)}= \text{dim}\mathbf{A}_{(2)}-\text{max}\left(\text{dim}\mathbf{A}_{(1)}^{\text{ub}}\right).
\end{equation}
In other words, the above relation defines unbroken and emergent symmetries as duals of each other in some cases. The exact settings realizing the above equation are worth further study.

For the readers interested in the perturbative quantum field theoretic description of our method, we note two possible research directions. The first is a straightforward analysis of the interpolation behavior of $H_{(1)}$ to $H_{(2)}$. This interpolation can be implemented by preparing the Hamiltonian $(1-\lambda)H_{(1)}+\lambda H_{(2)}$ with $\lambda=0$ to $\lambda=1$ and the perturbative analysis around $\lambda =0$ is useful also in QFTs. In this research direction, numerical methods such as the truncated conformal space approach\cite{Yurov:1989yu,Yurov:1991my} and analysis of corresponding lattice models will be fundamentally important. 

The second is the detection of a maximal unbroken subalgebra in this subsection. By assuming that a general perturbation $H_{\text{per}}$ from a UV CFT (possibly, multiple perturbations in QFTs) is invariant under the symmetry corresponding to this subalgebra $\mathbf{A}_{(1)}^{\text{ub}}$, one can restrict the form of possible perturbations systematically. More conventionally, one can write down the condition as follows,
\begin{equation}
[H_{\text{per}}, Q_{a'_{(1)}}]=0,
\end{equation}
where $Q_{a'_{(1)}}$ is a symmetry operator (or integral of motions) corresponding to the element $a'_{(1)} \in \mathbf{A}_{(1)}^{\text{ub}}$. Inversely, if the perturbation $H_{\text{per}}$ is given, one can determine possible RG flows by studying the maximal subalgebra structure which commutes with $H_{\text{per}}$ and its disjoint ideals.
Related arguments based on group symmetry can be seen ubiquitously, and our argument can be considered as a generalization emphasizing the ideal structure. We note \cite{Buican:2017rxc,Kikuchi:2021qxz,Kikuchi:2022gfi,Nakayama:2024msv,Kikuchi:2024cjd,Chen:2025qub,Gaberdiel:2026sfg,Ambrosino:2026umb} as references containing analysis of symmetry with ring structure and their action on bulk fields. 

For the readers interested in higher-dimensional systems, we note that the above arguments themselves can be applied straightforwardly to $D$-dimensional systems when treating the generalized symmetries as integral of motion or symmetry operators. When the IR theory becomes an extended TQFT under massive perturbations, the subring structure formed by $d$-dimensional symmetry operators in a UV theory will flow to that of the IR theory, where $d=0, ..., D$ specifies dimensionality of symmetry. This ring structure will be obtained from the application of the topological Wick rotation to composite defects in \cite{Shimamori:2024yms}. If the corresponding flows are massless, the quotient ring of UV theory appears as IR symmetry.

\subsection{Historical remarks on the projections in physics}

Before moving to the next section, we comment on the history of the ideas to represent RG as a projection. The idea itself can be seen ubiquitously in the fields, but well-known earlier examples related to the discussions in this work are the corner-transfer method by Baxter \cite{BAXTER198118}(for more details, see the review\cite{Baxter_2007}) and projective representation in Affleck-Kennedy-Lieb-Tasaki chain\cite{Affleck:1987vf}. This kind of projection or projective representation played a significant role in studying many-body quantum systems both theoretically and numerically. Related historical aspects can be seen in the review \cite{Cirac:2020obd}. 

Related to the argument in CFT in this manuscript, we also note the restriction of Hilbert space in quantum group symmetric models\cite{Pasquier:1989kd,LeClair:1989wy,Felder:1991hw} which have a connection to the Dotsenko-Fateev Coulomb gas\cite{Dotsenko:1984nm,Dotsenko:1984ad} (or Feigin-Fuchs construction\cite{Feigin:1981st} based on Wakimoto free field representation\cite{Wakimoto:1986gf}) and the massless RG flow\cite{Zamolodchikov:1987ti,Zamolodchikov:1987jf,Zamolodchikov:1989hfa}. More recently, the RG domain wall, which is in the main scope of this work, has been introduced in \cite{Brunner:2007ur} and its general correspondence with existing coset CFTs has been proposed in \cite{Gaiotto:2012np}. Some of the properties of homomorphism, or the properties of symmetry-preserving domain wall, have already appeared in \cite{Gaiotto:2012np}, and it is remarkable that this pioneering work appeared before the introduction of generalized symmetry\cite{Gaiotto:2014kfa}. Later, more general cases have been studied in \cite{Stanishkov:2016pvi,Stanishkov:2016rgv,Poghosyan:2022ecv,Poghosyan:2022mfw,Poghosyan:2023brb} and more mathematical understanding has been developed in \cite{Klos:2019axh,Klos:2020upw,Klos:2021gab,Kikuchi:2022rco,Kikuchi:2022ipr,Zeev:2022cnv,Cogburn:2023xzw,Benedetti:2024utz}. The idea of RG domain wall has also been combined with the holography in \cite{Tang:2023chv,Gutperle:2024yiz} and related bound of entropy of conformal interface has been discussed in \cite{Karch:2024udk}. More recently, the correspondence between symmetry-preserving domain wall\cite{Lan:2014uaa,kong2015boundarybulkrelationtopologicalorders,Wan:2016php,Kong:2017hcw,Kaidi:2021gbs} and RG domain wall has been realized in \cite{Zeev:2022cnv,Fukusumi_2022_c,Zhao:2023wtg,Fukusumi:2024nho,Fukusumi:2024ejk}, but this is relatively less familiar in the fields.

On the other hand, by interpreting a measurement as a projection, the measurement-induced quantum phase transition has been introduced in \cite{PhysRevB.98.205136,PhysRevB.100.134306,Skinner:2018tjl}. We note several works discussing the CFT description of the measurement induced quantum phase transition\cite{Li:2020pcv,Fuji:2020wlw,Yamamoto:2021tne,Oshima:2022yrw,Murciano:2023nqr,Kumar:2023evr,Ashida:2023ziz,Hoshino:2024kxk,Liu:2024pbd}. However, as far as we know, its relation to RG domain wall or massless RG flow has been rarely studied.

\section{Generating ideal: Role of nonabelian anyon or noninvertible symmetry}
\label{generating_ideal}

In this section, we introduce the easiest mathematical procedure for generating or constructing an ideal from an algebra $\mathbf{A}_{(1)}$. The discussion is standard and established in mathematics, so the main readers in this section are researchers in the physics community unfamiliar with abstract algebra. The benefit of our algebraic construction is summarized as follows,

\begin{equation}
\begin{split}
&\text{ If the fusion ring of UV theory has been given,} \\
&\text{ one can produce the possible IR fusion rings.}
\end{split}
\end{equation}
One can perform the exact construction by classifying ideals of a UV theory and obtain (infinitely) many new series of unexplored homomorphisms. In other words, if the above statement does not hold, the connection between the RG domain wall and massless flow will be broken. Further investigation of more detailed properties of the IR theory, such as conditions for the locality or nonlocality and conditions to obtain the IR theory with positive or integer fusion coefficients, is an open problem. However, surprisingly, we note that the related problems of the projection for Yang-Lee CFT flows have already been investigated much earlier in \cite{Smirnov:1990vm}. The categorical analog of the ideal called the additive congruence relation by ideal, will play a role in the more categorical description of our method\footnote{We thank Jurgen Fuchs for teaching the terminology to us.}. Related to this research direction, we remark that the idea of an ideal appears in the studies of Hopf algebra and related TQFTs (see \cite{etingof2024briefintroductionquantumgroups,schweigert2014hopf}, for example). We also remark that in the studies on Hopf algebra, one can see the ideas to represent a RG as projection \cite{Kreimer:1997dp,Connes:1998qv,Connes:1999yr,Connes:2000fe} but their formalism is closer to Lagrangian formalism and its precise connection to RG domain wall is (reasonable but) unclear at this stage (see also the reviews\cite{Panzer:2012gp,naseer2015hopfalgebrarenormalizationbrief})

\begin{figure}[htbp]
\begin{center}
\includegraphics[width=0.5\textwidth]{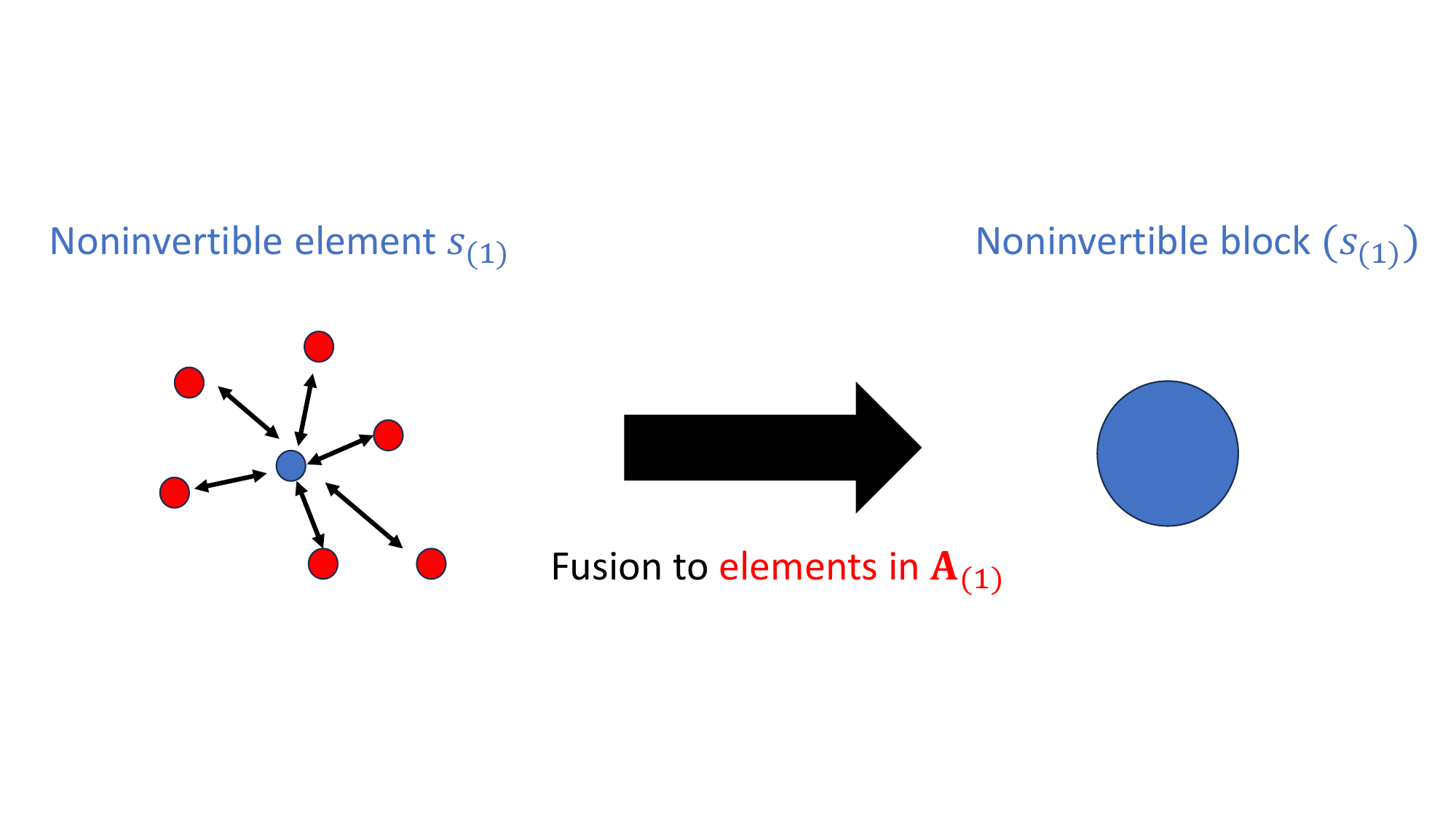}
\caption{Picture of the noninvertible object (anyon or symmetry) as a seed to generate a condensable block of a theory. We used blue color for noninvertible objects, black color for fusion between the noninvertible object and other objects, and red color for objects in $\mathbf{A}_{1}$. Because of the noninvertible structure, one can send the ideal generated by $s_{(1)}$ to $0$ only by applying a projection to the theory.}
\label{ideal_noninvertible}
\end{center}
\end{figure}

For simlpicity, we assume $\mathbf{A}_{(1)}$ is a commutative ring.
First, let us introduce an object $s_{(1)}\in \mathbf{A}_{(1)}$. Then the following set forms an ideal in $\mathbf{A}_{(1)}$,
\begin{equation}
(s_{(1)})=\{ s_{(1)}a_{(1)} \in  \mathbf{A}_{(1)} | a_{(1)}\in \mathbf{A}_{(1)} \}
\end{equation}
This is usually called an ideal generated by an element and denoted as $(s_{(1)})$. By introducing more elements represented as a set $S=\{ s_{(1),1}, ..., s_{(1),n}\}$ and considering their linear summation, one can obtain a more general ideal,
\begin{equation}
(S)=\left\{ \sum_{i=1}^{n}s_{(1),i}a_{(1),i} \in  \mathbf{A}_{(1)} | a_{(1),i}\in \mathbf{A}_{(1)}, i=1,..,n \right\}
\end{equation}
By considering the joint or intersection of such ideals, one can construct a series of ideals. This provides an intuitive and simple method of constructing homomorphisms.

In particular, if one takes a noninvertible object in $\mathbf{A}_{(1)}$, it generates a nontrivial ideal straightforwardly because the generated ideal cannot contain identity. From this view, one can say that a noninvertible symmetry corresponding to a nonabelian anyon is a seed to obtain an ideal and corresponding homomorphism or massless RG flow (FIG. \ref{ideal_noninvertible}). By combining this with Moore-Seiberg data\cite{Moore:1988qv,Moore:1988ss,Moore:1989vd,Fuchs:2002cm}, the fundamental importance of further systematic studies on noninvertible symmetries (or integral of motions) and the corresponding nonabelian anyons for the general understanding of TOs becomes more evident. 

One can construct possible IR theories only from the algebraic data of a UV theory. However, the construction of the homomorphisms between given UV and IR theories requires more involved calculations and consistency checks. In the next section, we solve this problem in several models.

\subsection{Emergent defects: intrinsically emergent objects}
In this subsection, we comment on the nontriviality of the defects appearing in the IR theory. As we have already demonstrated, the IR symmetry operators have been determined completely by the UV theory, under suitable assumptions. We have mainly discussed the transformation law of symmetry operators or anyons in higher dimensions and the coefficients before the objects can be non-integer by definition.

However, when studying the corresponding transformations of the defects, such an appearance of the non-integer coefficient is strictly prohibited. Hence, there can exist IR defects that cannot be related to the UV defects straightforwardly, by assuming the compatibility with the defect and bulk RG flows. Hence, we propose emergent defect (or emergent generalized symmetry in category theory) as follows,
\begin{equation}
\begin{split}
&\text{An IR defect $\mathcal{D}_{a_{(2)}}$ is emeregnt when there exists} \\
&\text{no UV defects that flow to it.}
\end{split}
\end{equation}

We note that the studies on the bulk and defect RG flow have still been limited. We note several related works on the bulk and boundary (or defect) RG flow \cite{Fredenhagen:2009tn,Dorey:2009vg,Chang:2018iay,Lauria:2023uca,Ambrosino:2025pjj}. Moreover, the anomaly inflow mechanism\cite{Callan:1984sa} complicates the systematic studies because one cannot apply the $g$-theorem straightforwardly to the bulk and defect RG process\cite{Green:2007wr}.

\section{Construction of homomorphism}
\label{construction_homomorphism}
In this section, we study ring homomorphisms between fusion rings appearing in massless RG flow of conformal field theories\cite{Zamolodchikov:1987ti,Zamolodchikov:1987jf,Zamolodchikov:1989hfa}. We study the massless RG flow between ultraviolet (UV) CFT $\mathbf{B}$ to infrared (IR) CFT $\mathbf{B}'$ where $\mathbf{B}$ and $\mathbf{B}'$ represent nonchiral fusion ring or anyon algebra of spherical fusion category (SFC)\cite{Rida:1999ru,Rida:1999xu,Nivesvivat:2025odb}. Roughly speaking, the SFC corresponds to the bosonic modular invariant and conformal bootstrap technique\cite{Polyakov:1974gs}. We concentrate our attention on the algebraic relation of objects or anyons, study the ring homomorphism $\rho$: $\mathbf{B}$ $\rightarrow$ $\mathbf{B}'$ (or algebraic data of a functor between two SFCs\cite{Kong:2013aya,kong2015boundarybulkrelationtopologicalorders,Kong:2017hcw}). To simplify the notation and discussion, we assume the theory is in the scope of Moore-Seiberg data\cite{Moore:1988qv,Moore:1988ss,Moore:1989vd,Fuchs:2002cm} and the ring homomorphism between nonchiral fusion rules naturally introduces a ring homomorphism between chiral fusion rules and vice versa. Algebraically, this is trivial because of the ring isomorphism between the nonchiral and chiral fusion rules. 

Before moving into the respective discussions, we remark on one of the most nontrivial points of our findings. Surprisingly, even in the flow from the tricritical Ising model to the Ising model, the homomorphism is not unique. However, by requiring a variant of integrability, the solution becomes unique. This phenomenon itself is analogous to ideal number decomposition by Kummer, and the integrability will play a fundamental role in constructing a treatable class of models. Inversely, the other homomorphisms are mysterious to some extent, but there exists a conjectural phenomenology to them. If one assumes the IR theory $\mathbf{A}_{(2)}$ or consequence of RG flow is integrable, one can say this homomorphism induces partially integrable embedding of the theory to UV theory $\mathbf{A}_{(1)}$, $\mathbf{A}_{(2)}\subset \mathbf{A}_{(1)}$. Fortunately, there exist very similar phenomena called partially solvable systems in recent literature\cite{Matsui_2024,Matsui2024BoundaryDS,Katsura:2024lrn}\footnote{We thank Chihiro Matsui for introducing us to the idea of this embedding and related ideas in the literature and Hosho Katsura for indicating useful references in this research field.}. Historical aspects of those models have been summarized in the introduction of \cite{Katsura:2024lrn}, and we note the Hilbert space fragmentation as a fundamental aspect, and note a recent review \cite{Moudgalya:2021xlu}. 

We conjecture that a ring homomorphism generally corresponds to integrable or partially solvable systems. Further analysis of the RG flows by perturbative QFT methods is an interesting direction, and it will be described by UV CFT with multiple perturbations as in \cite{Konechny:2023xvo}. We stress that such research direction is still in progress because of the difficulty of treating multiple perturbations straightforwardly (See the discussions in \cite{Gukov:2015qea,Gukov:2016tnp}), and our formalism will provide useful predictions as an alternative approach based on the quantum Hamiltonian.

\subsection{Tricritical Ising to Ising}

The tricritical Ising CFT has six primary fields, and one can decompose it into the tensor product of Ising and Fibonacci fusion rules. Hence, we denote the model as,
\begin{equation}
\{ I, \psi, \sigma \}\otimes \{ I, \tau \}
\end{equation}
where $\{ I, \psi, \sigma \}$ satisfy the following Ising fusion rule,
\begin{align}
\psi \times \psi&=I, \\
\psi \times \sigma &=\sigma, \\
\sigma \times \sigma &= I + \psi
\end{align}
and $\{ I, \tau \}$ satisfy the Fibonacci fusion rule,
\begin{equation}
\tau \times \tau =I +\tau
\end{equation}
In both fusion rules, $I$ represents the corresponding identity object. The remarkable point of this representation is the $Z_{2}$ group-like action of $\psi$, and this $Z_{2}$ structure is called a simple current in physics. Corresponding to this $Z_{2}$ like property, one can introduce the $Z_{2}$ charge. In this manuscript, we study $Z_{2}$ charge-preserving homomorphism or RG flow satisfying the t'Hooft anomaly matching condition. In the above fusion rule, $\{I, \psi\}$ are uncharged and $\{ \sigma\}$ is charged.

It is widely known that this CFT flows to the Ising CFT. However, as far as we know, the literature lacks the respective algebraic analysis. To distinguish the UV and IR theory, we denote the IR Ising CFT as $\{ I', \psi', \sigma' \}$. The conformal dimension of the objects are $h_{I'}=0, h_{\psi'}=1/2, h_{\sigma'}=1/16$.

To simplify the analysis, we assume the structure of the Ising fusion rule is preserved under the homomorphism,
\begin{align}
\rho(I)&=I', \\
\rho(\psi)&=\psi', \\
\rho(\sigma)&=\sigma'.
\end{align}
In this setting, one can construct a ring homomorphism when the homomorphism acts on the object $\tau$ canonically. Because of the theft anomaly matching condition, the action of $\rho$ to $\tau$ should be in the following form
\begin{equation}
\rho (\tau)=\alpha I' + \beta \psi'
\end{equation}
where $\alpha$ and $\beta$ are some complex numbers.

Because the object $\tau$ satisfies only one nontrivial fusion rule, the consistency condition of the fusion rule is simple,
\begin{equation}
\rho(\tau)\times \rho(\tau) =\rho(I) +\rho(\tau)
\end{equation}

Hence, by inserting the condition coming from the t'Hooft anomaly matching condition, one can obtain the following equations,
\begin{align}
\alpha^{2}+\beta^{2}&=\alpha+1 \\
2\alpha \beta =\beta
\end{align}

We obtain the following four solutions,
\begin{equation}
(\alpha, \beta)=\left(\frac{1}{2}, \pm \frac{\sqrt{5}}{2}\right) , \left(\frac{1\pm \sqrt{5}}{2},0 \right)
\end{equation}

It might be surprising for the readers in theoretical physics that only the preservation of symmetry coming from the symmetry-preserving domain wall condition cannot fix the homomorphism. We further introduce the assumption on the preservation of the fermion parity structure. In this setting, one can restrict the solutions to those with $\beta=0$. We can obtain a single solution by imposing the positivity of the coefficient. The solution corresponding to the RG domain wall is,
\begin{equation}
(\alpha, \beta)=\left(\frac{1+ \sqrt{5}}{2},0 \right)
\end{equation}

The other solutions can be interpreted as domain walls with domain wall particles when considering the folding trick. More precisely, the interpretation of these solutions in the context of massless RG is an interesting future problem. We conjecture that they are described by multiple terms of perturbations\cite{Konechny:2023xvo} and correspond to the partially solvable models in \cite{Matsui_2024,Matsui2024BoundaryDS,Katsura:2024lrn}.

Before going to the next example, we remark that we have assumed that $\psi$ is mapped to $\psi'$, but one can do the same analysis by replacing the chirality of $\psi$ and $\psi'$. 
Corresponding to the combinations of chiral or antichiral extension, there exists a series of homomorphisms. In condensed matter theory, this appears as particle-hole conjugate symmetry\cite{Son:2015xqa,Barkeshli:2015afa} and the above discussion provides an evident algebraic expression. Corresponding phenomena happen in general $Z_{2}$ symmetric models, including other examples in this work.

\subsection{$SU(2)_{1}\times SU(2)_{1} \rightarrow \text{Majorana}$}

Here, we study a simple model that plays a significant role in the study of SPT in condensed matter. We strudy the ring iosomorphism between $SU(2)_{1} \times SU(2)_{1}$ to $Z_{2}$ extension of $SU(2)_{2}$. We denote the chiral objects in $SU(2)_{1} \times SU(2)_{1}$ as,
\begin{equation}
\{ I^{[1]},  j^{[1]}\}\otimes \{ I^{[2]}, j^{[2]}\}
\end{equation}
They satisfy the double semion algebra. We regard $j^{[1]}j^{[2]}$ as $Z_{2}$ simple current and classify $\{ I,  j^{[1]}j^{[2]}\}$ as the uncharged sector and $\{ j^{[1]},j^{[2]}\}$ as charged sector.

Because $SU(2)_{2}$ satisfy the Ising fusion rule we denote the theory as $\{ I, \psi, \sigma \}$. It has been well established that the $Z_{2}$ extension of this chiral Ising fusion rule forms a double semion algebra\cite{Ginsparg:1988ui}. 

When considering the correspoding ring homomorphism between bulk $SU(2)_{1} \times SU(2)_{1}$ to bulk $SU(2)_{2}$, more detailed calculations are necessary. For this purpose, we introduce the following set of objects in $SU(2)_{1} \times SU(2)_{1}$,
\begin{equation}
\{ I^{[1]},  J^{[1]}\}\otimes \{ I^{[2]}, J^{[2]}\}
\end{equation}
where one can represent $J^{[i]}=j^{[i]}\overline{j^{[i]}}$. To avoid complications, we drop the upper indices of the identity operators. 

There exists no ring homomorphism in the bosonic representation, straightforwardly. However, one can consider the following unusual homomorphism,

\begin{align} 
\rho(I)&=I' \\
\rho(J^{[1]}J^{[2]})&=\epsilon', \\
\rho(J^{[1]})&=\frac{\sigma'_{\text{Bulk},+}+\sigma'_{\text{Bulk},-}}{\sqrt{2}}, \\
\rho(J^{[2]})&=\frac{\sigma'_{\text{Bulk},+}-\sigma'_{\text{Bulk},-}}{\sqrt{2}},
\end{align}
where we have modified the notation in the works by the first atuhor\cite{Fukusumi:2024cnl}. The nontrivial fusion rules of the IR theory are summarized as follows,
\begin{align}
\epsilon' \times  \epsilon'&=I', \\
\epsilon \times  \sigma'_{\text{Bulk},+}&=\sigma'_{\text{Bulk},+}, \\
\epsilon \times  \sigma'_{\text{Bulk},-}&=-\sigma'_{\text{Bulk},-}, \\
\sigma'_{\text{Bulk},+} \times  \sigma'_{\text{Bulk},+}&=I'+\epsilon', \\
\sigma'_{\text{Bulk},-} \times  \sigma'_{\text{Bulk},-}&=I'-\epsilon', \\
\sigma'_{\text{Bulk},+} \times  \sigma'_{\text{Bulk},-}&=0, 
\end{align}

It should be stressed that $\sigma'_{\text{Bulk},+}\times\sigma'_{\text{Bulk},-}$ is vanishing and one needs to introduce two sets of models when considering its lattice realization\cite{Yang_2004,Hagendorf_2012,Hagendorf:2012fz,Matsui:2016oqq}. This structure has appeared in fermionic string theory, for example, in \cite{Ginsparg:1988ui}, but it has never captured sufficient attention in the fields. This suggests the flows of the sectors $J^{[1]}$ or $J^{[2]}$ do not behave well formally, and the $Z_{2}$ symmetric or $Z_{2}$ antisymmetric sectors $(J^{[1]}\pm J^{[2]})/\sqrt{2}$ are more useful. The corresponding structure can be seen when studying the subalgebra of $SU(2)_{1}\times SU(2)_{1}$.

One can also obtain the $Z_{2}$ extended version of the homomorphism by applying the chiral $Z_{2}$ extension of the left and right-hand side of the homomorphism.

\begin{align} 
\rho(j^{[1]}j^{[2]})&=\psi' \\
\rho(\overline{j^{[1]}}\overline{j^{[2]}})&=\overline{\psi'}, \\
\rho (\overline{j^{[1]}} j^{[2]})&=\frac{\mu'_{\text{Bulk},+}+\mu'_{\text{Bulk},-}}{\sqrt{2}}, \\
\rho(j^{[1]}\overline{j^{[2]}})&=\frac{\mu'_{\text{Bulk},+}-\mu'_{\text{Bulk},-}}{\sqrt{2}},
\end{align}
where $\mu'_{\text{Bulk},+}$ and $\mu'_{\text{Bulk},-}$ are disorder operator and they are introduced by considering the $Z_{2}$ extension $\mu'_{\text{Bulk},+}=\psi' \times \sigma'_{\text{Bulk},+}$ $\mu'_{\text{Bulk},-}=\psi' \times \sigma'_{\text{Bulk},-}$.

\subsection{$\{SU(2)_1\}^3 \to SU(2)_3$}

Here we study the homomorphism induced from the flow $\{SU(2)_1\}^3 \to SU(2)_3$.
We use the same notation in the previous subsection, and we distinguish the simple currents of each $SU(2)_1$ by writing them as $J^{[i]}$, $i=1,2,3$.
In $SU(2)_3$ there are four primary fields with spins ranging from $0$ to $\frac{3}{2}$, which we list in ascending order of spin as ${I',x',\tau',J’}$.
Here, $\tau'$ is the primary field that satisfies the Fibonacci fusion rule $\tau'\times\tau' =1+\tau'$, and $J’$ is the simple current of $SU(2)_3$. Here we study the homomorphism preserving $Z_{2}$ structures $\{I, J^{[1]}J^{[2]}J^{[3]}\}$ and $\{ I, J'\}$.

Because each theory has a $\mathbb{Z}_2$ symmetry generated by its simple current, we can write the relations,
\begin{equation}
    \rho(J^{[i]}J^{[j]})=\alpha^{ij}I'+\beta^{ij}\tau' ,\quad i\neq j,\alpha^{ij},\beta^{ij}\in\mathbb{C} 
\end{equation}
and regard this as a suitable linear combination. In this representation, we have assumed two conditions, preservation of the $Z_{2}$ group ring structure and t'Hooft anomaly matching condition as in the previous cases. 
To determine the specific values of $\alpha^{ij}$ and $\beta^{ij}$, we use the fusion rules among the $J^{[i]}$.
Specifically, we exploit $I=I\times I=(J^{[i]})^{2} (J^{[j]})^{2}=(J^{[i]}J^{[j]})^{2}$.
Together with the previous equation and nontrivial fusion rule $\tau' \times \tau'=I'+\tau'$ results in the following form,
\begin{equation}
    1=(\alpha^{ij}1+\beta^{ij}\tau')^2=((\alpha^{ij})^2+(\beta^{ij})^2)1 + (2\alpha^{ij}\beta^{ij}+(\beta^{ij})^2)\tau'
\end{equation}

Hence, for any $i,j,(i\neq j)$,
\begin{equation}
    (\alpha^{ij})^2+(\beta^{ij})^2=1,\quad \beta^{ij}(2\alpha^{ij}+\beta^{ij})=0
\end{equation}
is obtained.
Thus we find two possible solutions $(\alpha^{ij},\beta^{ij})=(1,0)$ and $\bigl(\tfrac{1}{\sqrt{5}},-\tfrac{2}{\sqrt{5}}\bigr)$; the full set of $(\alpha^{ij},\beta^{ij})$ for $i,j\in{1,2,3},,i\neq j$ can be obtained as follows.

Let $i,j,k$ be mutually distinct labels.
Since $J^{(i)}J^{(j)}=J^{(i)}(J^{(k)})^2 J^{(j)}=(J^{(i)}J^{(k)})(J^{(k)}J^{(j)})$, we have $\alpha^{ij}1+\beta^{ij}\tau=(\alpha^{ik}1+\beta^{ik}\tau')(\alpha^{kj}1+\beta^{kj}\tau')=(\alpha^{ik}\alpha^{kj}+\beta^{ik}\beta'^{kj})1+(\alpha^{ik}\beta^{kj}+\alpha^{kj}\beta^{ik}+\beta^{ik}\beta^{kj})\tau'$.
If we assume $\beta^{ij}\neq 0$, then $\beta^{ij}=-\tfrac{2}{\sqrt{5}}$, so
\begin{equation}\label{eq:4}
    \alpha^{ik}\beta^{kj}+\alpha^{kj}\beta^{ik}+\beta^{ik}\beta^{kj}=-\frac{2}{\sqrt{5}}
\end{equation}
holds.
However, if $\beta^{ik}\beta^{kj}\neq 0$, we would have to set $\beta^{ik}\beta^{kj}=\tfrac45$, which cannot satisfy equation \eqref{eq:4}.
Hence, without loss of generality we set $\beta^{ik}=0$, which gives $\beta^{kj}=-\tfrac{2}{\sqrt{5}}$ and $\alpha^{kj}=\tfrac{1}{\sqrt{5}}$.
Using $J^{[i]}J^{[k]}=(J^{[i]}J^{[j]})(J^{[j]}J^{
k]})$ we then have
\begin{equation}
    I'=(\frac{1}{\sqrt{5}}I'-\frac{2}{\sqrt{5}}\tau')(\frac{1}{\sqrt{5}}I'-\frac{2}{\sqrt{5}}\tau')
\end{equation}
which is consistent. We note that contrary to the naive intuition from the fact that $\tau'$ is a nonabelian anyon, above $I'/\sqrt{5}-2\tau/\sqrt{5}$ is invertible. One can also check the invertibility of $\tau'$ itself by calculating $\tau'\times(\tau'-I')=I'$.
In summary, the admissible solutions are either $(\alpha^{ij},\beta^{ij})=(1,0)$ for every $i\neq j$, or the pattern in which two out of the three unordered pairs $(i,j)$ take $\bigl(\tfrac{1}{\sqrt{5}},-\tfrac{2}{\sqrt{5}}\bigr)$ while the remaining pair takes $(1,0)$. It is remarkable that when assuming the homomorphism is surjective, the appearance of the fractional and negative coefficients is inevitable in this case. When one prohibits the appearance of such an unusual coefficient, the homomorphism is isomorphic to the homomorphism $\{ SU(2)_{1}\}^{3} \rightarrow SU(2)_{1}$ and this is consistent with the observation in \cite{Furuya:2015coa}.

\subsection{$\{SU(2)_{1}\}^4 \to SU(2)_4$: puzzle of charged sector}
Next, we determine the homomorphism corresponding to the flow $\{SU(2)_{1}\}^{4}$ $\to$ $SU(2)_4$ in uncharged $Z_{2}$ sector. We keep the notation for $SU(2)_1$ as before and label each copy by the upper indices $[i]$ and take the $Z_{2}$ simple current as $J=\prod_{i=1}^{4}J^{[i]}$. The theory $SU(2)_4$ has five chiral primary fields, labeled by their spin as $\phi'_0,\ \phi'_{1/2},\ \phi'_{1},\ \phi'_{3/2},\ \phi'_{2}$  whose fusion rules are listed in the table below. The remarkable point of these two theories is the conformal dimension $1$ of the $Z_{2}$ currents $J$ and $J'$, and this means that the theories are classified as the simplest bosonic model in the corresponding series of models.
\begin{table}[htbp]
  \centering
  \caption{Fusion rules of SU(2)\(_4\) WZW primaries}
  \renewcommand{\arraystretch}{1.2} 
  \begin{tabular}{c|ccccc}
    $j_1 \!\backslash\! j_2$ & $0$ & $1/2 $ & $1$ & $3/2$ & $2$ \\ \hline
    $0$            & $0$                         & $1/2$                     & $1$                         & $3/2$                     & $2$ \\ 
    $\dfrac12$     &                              & $0 + 1$                        & $(1/2) + (3/2)$       & $1 + 2$                        & $3/2$ \\ 
    $1$            &                              &                                 & $0 + 1 + 2$                 & $(1/2) + (3/2)$          & $1$ \\ 
    $3/2$     &                              &                                 &                              & $0 + 1$                        & $1/2$ \\ 
    $2$            &                              &                                 &                              &                                 & $0$
  \end{tabular}
\end{table}
$\phi_2(=J’)$ is the $\mathbb{Z}_2$ simple current of the theory. By computing the $Z_{2}$ charge of each operator concerning $J$ or $J’$, we can restrict the operators that should be connected by the homomorphism. In the present case, $\phi_{1/2}$ and $\phi_{3/2}$ belong to the $Z_{2}$ uncharged sector and the others belong to the $Z_{2}$ charged sector. 
 One must note, as in the Majorana-fermion case, that $Z_{2}$ invariant operators split after the $Z_{2}$ simple current extension. In the present case the $\mathbb{Z}_2$-invariant sector is $\phi'_{1}$, which therefore splits into $\phi'_{1,0}$ and $\phi'_{1,1}$ where the lower index $0$, $1$ corresponds to $Z_{2}$ fermion parity. Even after the $Z_{2}$ extension, the number of operators in the uncharged and charged sectors is significantly different, and this is in sharp contrast to the Majorana fermion case with the fermionic T-duality (or a variant of supersymmetry in older literature\cite{Gliozzi:1976qd}). Moreover, in this case, the $Z_{2}$ invariant sector is uncharged.

 We must first determine the fusion rules that include these operators, especially uncharged operators for ${\phi'_{0},\phi'_{1,0},\phi'_{1,1},\phi'_{2}}$. As explained in paper \cite{Fukusumi:2024cnl}, these are obtained by taking a $\mathbb{Z}_2$ extension of the original fusion-category data and then extracting its subring. Related discussions can be seen in the recent works\cite{Huang:2023pyk,Wen:2024udn,Huang:2024ror,Bhardwaj:2024ydc}, and one can see its categorical interpretations. First, we determine the nonchiral fusion ring from the Moore-Seiberg data\cite{Moore:1988qv,Moore:1988ss,Moore:1989vd,Fuchs:2002cm}. The procedure is relatively simple,
\begin{equation}
\{ \phi'_{0}, \phi'_{1/2}, \phi'_{3/2}, \phi'_{2}, \phi'_{1} \} = \{ \Phi'_{0}, \Phi'_{1/2}, \Phi'_{3/2}, \Phi'_{2}, \Phi'_{1,0} \}
\end{equation}
where $=$ is ring isomorphism and $\Phi'$ is the objects or anyons in nonchiral fusion ring\cite{Rida:1999xu,Rida:1999ru,Nivesvivat:2025odb} (or spherical fusion category). For a conventional reason, we changed the lower index of $\phi'_{1}$ to $1,0$ in $\Phi'_{1,0}$. The $Z_{2}$ extension can be achieved by multiplying the chiral $Z_{2}$ structure to this fusion ring. For example, one can obtain $\phi'_{0}\overline{\phi'_{2}}=J'\Phi'_{0}$. Because of the extension, it is necessary to introduce the nontrivial object $\Phi'_{1,1}=J'\Phi'_{1,0}$, and this corresponds to a disorder operator.

By considering the anyon condensation process, we can obtain the algebraic data of the symmetry topological field theory (SymTFT) $\{\Psi \}$\cite{Apruzzi:2021nmk}, which is ring isomorphic to the extended chiral algebra. For simplicity, we concentrate on uncharged sectors which form a subalgebra and denote the condensation as follows,
\begin{align}
    \Psi'_{0}&=\frac{\Phi'_0+\Phi'_2}{2},\\ 
\Psi'_2 &=\frac{J'\Phi'_0 +J'\Phi'_2}{2},\\ 
\Psi'_{1,0}&=\frac{\Phi'_{1,0}}{\sqrt{2}},\\ 
\Psi'_{1,1}&=\frac{\Phi'_{1,1}}{\sqrt{2}}
\end{align}
In particular, the fusion rules become those in the table below, which prepares us to determine the ring homomorphism. 
\begin{table}[htbp]
  \centering
  \caption{Fusion rules of $\mathbb{Z}_2$ extended SU(2)\(_4\) WZW primaries}
  \renewcommand{\arraystretch}{1.2} 
  \begin{tabular}{c|cccc}
    $j_1 \!\backslash\! j_2$ & $\Psi'_0$ &  $\Psi'_{1,0}$ & $\Psi'_{1,1}$ & $\Psi'_2$ \\ \hline
    $\Psi'_0$                   & $\Psi'_{0}$      & $\Psi'_{1,0}$        & $\Psi'_{1,1}$    & $\Psi'_2$ \\ 
    $\Psi'_{1,0}$            &             & $\Psi'_0 + \frac{1}{\sqrt{2}}\Psi'_{1,0} $    & $\Psi'_2 +  \frac{1}{\sqrt{2}}\Psi'_{1,1}$    & $\frac{\Psi'_{1,0}+\Psi'_{1,1}}{2}$ \\ 
    $\Psi'_{1,1}$   &             &           & $\Psi'_0 +\frac{1}{\sqrt{2}}\Psi'_{1,0}$      & $\frac{\Psi'_{1,0}+\Psi'_{1,1}}{2}$ \\ 
    $\Psi'_2$            &              &               &           &      $\Psi'_0$ \\
  \end{tabular}
\end{table}

We replace the $\Psi'$ with the $\phi'$ because of the ring isomorphism (or the CFT/TQFT correspondence or CFT/SymTFT correspondence). By evaluating the charge of each operator with respect to $J’$, we find that a product of two $SU(2)_1$ simple currents such as $J'^{[i]}J'^{[j]}$ is mapped to a linear combination of $\phi'_0,\ \phi'_{1,0},\ \phi'_{1,1},\ \phi'_2$. Thus we set
\begin{equation}
    \rho(J^{[i]}J^{[j]})=\alpha^{ij}\phi'_0+\beta_0^{ij}\phi'_{1,0}+\beta_1^{ij}\phi'_{1,1}+\gamma^{ij}\phi'_2\quad(i\neq j)
\end{equation}
and seek the possible coefficient quadruples $(\alpha^{ij},\beta_0^{ij},\beta_1^{ij},\gamma^{ij})$.

Because every generator of $\mathbb{Z}_2$ squares to the identity,
\begin{equation}
    \begin{split}
        &(\alpha^{ij})^2+(\beta_0^{ij})^2+(\beta_1^{ij})^2+(\gamma^{ij})^2=1,\\    &2\alpha^{ij}\beta_0^{ij}+\beta_0^{ij}\gamma^{ij}+\beta_1^{ij}\gamma^{ij} + \frac{(\beta_0^{ij})^2+(\beta_1^{ij})^2}{\sqrt{2}} \\
&=2\alpha^{ij}\beta_1^{ij}+\beta_0^{ij}\gamma^{ij}+\beta_1^{ij}\gamma^{ij}+\frac{2\beta_0 \beta_1}{\sqrt{2}}\\
        &=\beta_0^{ij}\beta_1^{ij}+\alpha^{ij}\gamma^{ij}=0
    \end{split}
\end{equation}
must hold. We can find the solutions by dividing into four cases according to whether $\beta_0^{ij}$ and $\beta_1^{ij}$ are both zero, only one is zero, or neither is zero. The answers are the solutions in which exactly one of $\alpha^{ij},\gamma^{ij}$ is $\pm1$, and the ten non-trivial solutions
\begin{equation}
\begin{split}
        &\alpha^{ij}\phi'_0+\beta_0^{ij}\phi'_{1,0}+\beta_1^{ij}\phi'_{1,1}+\gamma^{ij}\phi'_2 \\
        &=
        \begin{cases}
            \pm\frac{1}{3}\phi'_0 \mp \frac{2\sqrt{2}}{3}\phi'_{1,0}\\
            -\frac{2}{3}\phi'_0 \pm\frac{\sqrt{2}}{3}(\phi'_{1,0} + \phi'_{1,1})+\frac{1}{3}\phi'_2 \\
            \frac{1}{3}\phi'_0 \pm\frac{\sqrt{2}}{3}(\phi'_{1,0} + \phi'_{1,1})-\frac{2}{3}\phi'_2 \\
            \mp\frac{1}{3}\phi'_0 \pm\frac{\sqrt{2}}{3}(\phi'_{1,0} - \phi'_{1,1})\mp\frac{2}{3}\phi'_2 \\
            \pm\frac{2}{9}\phi'_0 \pm\frac{\sqrt{2}}{9}(\phi'_{1,0} + 5\phi'_{1,1})\mp\frac{5}{9}\phi'_2 \\
        \end{cases}.        
\end{split}
\end{equation}
According to this consistency condition, the solutions $\pm\frac{1}{3}\phi_0 \mp \frac{2\sqrt{2}}{3}\phi_{1,0}$ and $\pm\frac{2}{9}\phi_0 \pm\frac{\sqrt{2}}{9}(\phi_{1,0} + 5\phi_{1,1})\mp\frac{5}{9}\phi_2$ are excluded. This is because multiplying this expression by any other solution cannot make the coefficient of $\phi_0$ vanish. Therefore, the essentially non-trivial solutions are these six in total.
\begin{equation}
\begin{split}
 &\alpha^{ij}\phi_0+\beta_0^{ij}\phi_{1,0}+\beta_1^{ij}\phi_{1,1}+\gamma^{ij}\phi_2 \\
        &=
        \begin{cases}
            -\frac{2}{3}\phi_0 \pm\frac{\sqrt{2}}{3}(\phi_{1,0} + \phi_{1,1})+\frac{1}{3}\phi_2 \\
            \frac{1}{3}\phi_0 \pm\frac{\sqrt{2}}{3}(\phi_{1,0} + \phi_{1,1})-\frac{2}{3}\phi_2 \\
            \mp\frac{1}{3}\phi_0 \pm\frac{\sqrt{2}}{3}(\phi_{1,0} - \phi_{1,1})\mp\frac{2}{3}\phi_2 \\
        \end{cases}.
\end{split}    
\end{equation}

Once, for example, the solution for $(i,j)=(1,2)$ is fixed, the remaining coefficients can be determined consistently from\footnote{Note that the sign choices involved in determining these other coefficients still allow some freedom, but one needs to pick from at most two options.}
\begin{equation}
    \rho(J^{[1]}J^{[2]}J^{[3]}J^{[4]})=\phi'_2,\quad \rho(J^{[1]}J^{[2]})\rho(J^{[2]}J^{[3]})=\rho(J^{[1]}J^{[3]}).
\end{equation}
Hence we established the connection between the $Z_{2}$ uncharged sectors of $\{SU(2)_{1} \}^{4}$ and $SU(2)_{4}$. We remark again that the $SU(2)_{4}$ WZW model is one of the simplest models with bosonic $Z_{2}$ simple current because the conformal dimension of the $Z_{2}$ simple current $J'$ is $1$. In other words, for further studies on homomorphisms between more general CFTs, general numerical methods are necessary.   

Surprisingly, it is impossible to extend the homomorphism to uncharged sectors because of the incompatibility of their fusion rules of the $Z_{2}$ group structure in the $\{ SU(2)_{1}\}^{4}$ and their nonabelian fusion of $SU(2)_{4}$. For example, by assuming the t'Hooft anomaly matching condition, one can assume the form as follows,
\begin{equation}
\rho (J^{[k]})=\delta^{k}\phi'_{1/2}+\epsilon^{k}\phi'_{3/2}
\end{equation}
However, because of the relation $\rho(J^{[k]}J^{[k]})=\rho(I)=\phi'_{0}$ and the fusion rule between $\phi'_{1/2}$ and $\phi'_{3/2}$, it is impossible to obtain a solution. Hence, regardless of the close connection between the $\{ SU(2)_{1}\}^{4}$ and $SU(2)_{4}$(see \cite{Quella:2019los}, for example), it is impossible to obtain a surjective ring homomorphism between them at this stage. We expect that this puzzle can be resolved by considering combinations of other nonanomalous $Z_{2}$ extensions of $\{ SU(2)_{1}\}^{4}$ and the corresponding noninvertible extension of $SU(2)_{4}$ but studies on such noninvertible extension are still under development. Hence, we leave this puzzle as an open problem. 

We also remark that by generalizing the arguments for the $SU(2)_{2}\rightarrow \text{Majorana}$ case, it might be possible to obtain the ring homomorphism by taking the subring of the UV theory containing the charged sector. For example, the charged sector formed by $2$ $2$-dimensional basis $(\sum_{i}J^{[i]})/4$ and $(\sum_{i}JJ^{[i]})/4$ will be a reasonable charged subsector constructing the UV subalgebra.

\section{conclusion}
\label{conclusion}

In this work, we established the explicit construction of a ring homomorphism between fusion rings that determines the possible IR symmetry systematically. Moreover, their implications for massive RG, or abstract algebraic representation of the sandwich constructions in recent literature, have been proposed and shown. We conclude that the fundamental structure of symmetry in physics, the group, has been generalized to a ring in the context of the RGs.

We note again that, even in the well-established models connected by massless RG flows, a nontrivial structure appears. To obtain the homomorphism more generally, systematic studies on the ideal structure of the model are necessary, as we have stressed in the section. \ref{substructure}. When the ideal is commutative, the existence of an eigenvalue $0$  is guaranteed, and its existence in lattice models seems promising, in principle (However, we note a more involved problem of emanant symmetry in \cite{Cheng:2022sgb,Seiberg:2023cdc,Seiberg:2024gek}). In other words, if the ideal is not commutative, this implies difficulty in obtaining the IR theory by projection, and the IR symmetry will become more difficult to realize in a lattice model without taking some limit. Further studies of the ideal structure will be fundamentally important for the classification and construction of general TOs. In this work, we have restricted our attention to fusion rings appearing in $2$ $2$-dimensional CFT or $2+1$ dimensional TQFT, but the same construction of RG by ideal should apply to higher-dimensional systems. From this view, nontrivial transformations of UV and IR symmetry, as in \cite{Seiberg:2025bqy,Gu:2025gtb}, are worth further study.

We also note a related research direction toward coupled models with extended symmetries. By using the folding trick\cite{Wong:1994np}, it is known that the construction of extended model $\mathbf{F}_{\text{UV}\times \text{IR}}$ with  $\mathbf{A}_{(1)}\otimes \mathbf{A}_{(2)}\subset \mathbf{F}_{\text{UV}\times \text{IR}}$ plays a fundamental role also for the construction of homomorphism\cite{Gaiotto:2012np}. In a $Z_{N}$ symmetric case,  the extension is constructed by studying a common subgroup of $\mathbf{A}_{(1)}$ and $\mathbf{A}_{(2)}$. The anomaly-free object in these theories is the integer spin simple current\cite{Schellekens:1989uf,Schellekens:1990ys,Gato-Rivera:1990lxi,Gato-Rivera:1991bcq,Gato-Rivera:1991bqv,Kreuzer:1993tf,Fuchs:1996dd,Fuchs:1996rq} and its nonchiral generalizations\cite{Kong:2019cuu,Fukusumi:2024ejk} and the existence of this structure enables one to obtain the extended theory $\mathbf{F}_{\text{UV}\times \text{IR}}$. Hence, one expects the same strategy for the common subalgebra structure, but the gauging or extension for the product of this common subalgebra has not been studied as far as we know. As a natural extension of the $Z_{N}$ case, the nonsimple current in the coupled model forms a nonsimple current with (half) integer spin after taking the Deligne product (This condition corresponds to single-valuedness of the wavefunctions in TOs. The term nonsimple current appeared in the study of heterotic string theories. This indicates nontrivial extension or insertion of duality defects as in\cite{Petkova:2000ip,Frohlich:2004ef,Frohlich:2006ch}). For this purpose, recent progress on the orbifolding procedure for the noninvertible symmetry \cite{Diatlyk:2023fwf,Lu:2025gpt} may be useful. However, we stress that it is necessary to invent the corresponding extension, which is the inverse operation of the orbifolding. After the extension, the representation of the system will become intrinsically anyonic\cite{Feiguin:2006ydp,Buican:2017rxc}.

Finally, we comment on the interpretation of projection in the manuscript based on more recent literature on open quantum systems.
In this manuscript, we have restricted our attention to static quantum phases and their transition, but by interpreting this projection as a measurement, our construction of the projection will produce a kind of treatable measurement-induced quantum phase transition\cite{PhysRevB.98.205136,PhysRevB.100.134306,Skinner:2018tjl} satisfying $c$ and $g$-theorem\cite{Zamolodchikov:1986gt,Affleck:1991tk,Friedan:2003yc}. Further studies on the interpretation of the ideal structure in a quantum many-body system and its realization in an experimental setting are interesting research directions. Moreover, by interpreting the ring with a negative or fractional coefficient as a category, premodular fusion categories will appear, and they are expected to describe TOs in mixed states\cite{Sohal:2024qvq,Ellison:2024svg,Kikuchi:2024ibt}.

\section{Acknowledgement}
The first author thanks Guangyue Ji, Ha-Quang Trung, and Bo Yang for the interesting discussions on the hierarchical structure of fractional quantum Hall systems and Taishi Kawamoto for the collaborations related to this project. The first author also thanks Takamasa Ando, Weiguang Cao, and Pochung Chen for interesting discussions on noninvertible symmetries, Takumi Fukushima for sharing his knowledge of fracton topological orders, and Yifan Liu for reminding us of the significance of representing RG as a projection in a different research project, and Chihiro Matsui and Hosho Katsura for sharing their knowledge on the partial solvable systems and for the related discussions. The first author thanks Sylvain Ribault for many useful comments on the earlier version of the draft. The first author thanks Gerard Watts for pointing out the correspondence between the nontrivial homomorphism from the tricritical Ising CFT to the Ising CFT and phase diagram in \cite{Konechny:2023xvo} and Jurgen Fuchs for notifying us of terminology in category theory and for related useful comments. We also thank Yuya Kusuki for a useful comment. The first author thanks the support from NCTS and CTC.

\clearpage

\appendix

\section{Adding integral of motions to Hamiltonian or inducing gauge fixing}
\label{ideal_gauge_fixing}

In this section, we propose an alternative method for representing RG by implementing an analogy to gauge fixing in lattice gauge theory and the quotient by an ideal. As in the main text, we assume the UV Hamiltonian $H_{(1)}$ has $\mathbf{A}_{(1)}$ symmetry and it can be deformed to Hamiltonian $H_{(2)}$ with quotient ring $\mathbf{A}_{(1)}/\text{Ker} \rho =\mathbf{A}_{(2)}$ symmetry. Then we take the linear basis of  $\text{Ker} \rho$ as $\{ s_{(1),i}\}_{i=1}^{n}$ and consider deformations of the Hamiltonian $H_{(1)}$ under some terms represented as $Q_{s_{(1),i}}$ where $Q$ is matrix representations of the ideal.

Under the RG flow $\rho$, $\text{Ker} \rho$ can be sent to zero, and this is the most fundamental phenomenology in the main text. In other words, if this condition $\text{Ker} \rho=0$ can be enforced by adding some terms to the Hamiltonian $H_{(1)}$, one can expect the deformed Hamiltonian to capture the properties of the Hamiltonian $H_{(2)}$. In the main text, we interpret the condition $\text{Ker} \rho=0$ as mass or anyon condensation, but one can interpret this condition as another familiar phenomenon in physics. This can be interpreted as a gauge fixing.

First, we introduce the following deformed Hamiltonian,
\begin{equation}
H_{(1\rightarrow 2)}(\{ g_{i}\})=H_{(1)}+\sum _{i=1}^{n} g_{i}Q_{s_{(1),i}}Q_{s_{(1),i}}^{\dagger}
\label{ideal_gauge}
\end{equation}
where the constants $\{ g_{i}\}_{i=1}^{n}$ are coupling constants and $\dagger$ is the Hermitian conjugate. By taking all coupling constant $\{ g_{i}\}_{i=1}^{n}$ sufficiently large positive number, the energy splitting between the sector $\text{Ker} \rho=0$ and the other sectors become large, and one can obtain the Hamiltonian $H_{(2)}$ as low energy effective Hamiltonian,
\begin{equation}
H_{(1\rightarrow 2)}(\{ g_{i}\rightarrow \infty \})\sim H_{(2)}
\end{equation}
A similar analysis for the group generator can be seen in the previous work by the first author (with a close connection to Jordan-Wigner transformation)\cite{Fukusumi_2022_c}, but this case is significantly different from the case in \cite{Fukusumi_2022_c}, because the degree of freedom (typically central charge) is reduced under the procedure. 
 When the flow is integrable, this construction implies the existence of separated integrable sectors at high energy, and a similar structure has been reported in \cite{Shibata:2019vsf}. 

By taking respective high energy sectors labeled by $Q_{s_{(1),i}}=q_{(1),i}$ where $q_{(1), i}$ is an eigenvalue and there exists $i$ with $q_{(1), i}\neq 0$, one may expect to obtain a series of deformed ring structures $\mathbf{A}_{(1), \{s_{(1),i}=q_{i}\}}$ for each sectors. However, because of the property of $a_{(1)}\times \text{Ker} \rho=\text{Ker} \rho$ for all $a_{(1)}\in \mathbf{A}_{(1)}$, the structure $\{ q_{(1), i} \}$ completely determines all $a_{(1)}\in \mathbf{A}_{(1)}$ as a scalar. In other words, only the sector $\text{Ker} \rho =0$ generates a nontrivial emergent ring structure or symmetry denoted as $\mathbf{A}_{(1)}/\text{Ker}\rho$ when fixing or condensing the structure $\text{Ker} \rho$. However, by taking a subrng of $\mathbf{R}=\{ r_{i}\}_{i=1}^{m}\subset \text{Ker} \rho$ with $m<n$, one can still obtain a nontrivial deformed ring when the subrng $\mathbf{R}$ does not form an ideal in $\mathbf{A}_{1}$. Hence, we propose the following structure as a deformed ring (or ring after projection),
\begin{equation}
\mathbf{A}_{(1), \{r_{(1),i}=q_{i}\}}, \ \text{with} \ \mathbf{R}=\{ r_{i}\}_{i=1}^{m}\subset \text{Ker} \rho
\end{equation}
The value $q_{i}$ corresponds to the quantum dimension of the objects as a consequence of homomorphism between $\mathbf{A}_{(1)}\rightarrow \mathbf{C}$ where $\mathbf{C}$ is the complex number fields, but this will require more detailed investigations.

We also note the analogy between Eq. \eqref{ideal_gauge} and the effective Hamiltonian appearing in generalized Gibbs ensemble\cite{PhysRevLett.98.050405,PhysRevA.74.053616} which has a close connection to the quantum Korteweg–De Vries equation\cite{Bazhanov:1994ft} (For more general aspects, see discussions in \cite{Fukusumi:2024nho}). In other words, the perturbations added to $H_{(1)}$ induce thermalization, and one can interprete RG and resultant $c$- or $g$-theorem as a law of thermodynamics\cite{Zamolodchikov:1986gt,Affleck:1991tk,Friedan:2003yc}. To some extent, this is a standard phenomenological understanding of RG, but one can see the analogy more evidently by observing the structures coming from the ideal.

\section{Conjecture on coset representation of symmetry preserving domain walls}
\label{coset_junction}

Generalizing the argument in existing literature, we conjecture the decomposition of $UV$ theories by coset CFTs\cite{Goddard:1984vk,Goddard:1984hg,Goddard:1986ee,Goddard:1988md} or that by level-rank dualities\cite{Kuniba:1990im,Kuniba:1990zh,Nakanishi:1990hj}. The symbol ``$/$" in this section is different from that of the quotient ring in the main text, so we placed the arguments here. As far as we have checked, the construction of ideals in the main text has not been achieved in general, and the study of the possible homomorphism between theories described by coset CFTs will be useful. The correspondence between coset conformal field theory and massless RG flow has been studied in \cite{LeClair:2001yp}, and its connection to current-current deformation has been studied in the context of $\lambda$ deformation\cite{Sfetsos:2017sep,Georgiou:2017jfi,Georgiou:2018gpe,Borsato:2023dis}. Related quantum wire junction problems have been studied in \cite{Quella:2006de,Kimura:2014hva,Kimura:2015nka}. We also note more recent literature from different research motivations\cite{Bourgine:2024ycr,Cordova:2025eim,Antinucci:2025uvj}. In this section, we denote chiral or antichiral CFT as $M$ or $\overline{M}$ and label them by lower index IR or UV to identify them as UV and IR theories, respectively.

In $Z_{N}$ symmetric models, there exist two types of $Z_{N}$ symmetry preserving RG flow, anomaly matching flow, and anomaly cancellation flow\cite{Fukusumi_2022_c,Fukusumi:2024ejk}. For example, in the flow between $SU(N)_{k}$ WZW model to $SU(N)_{k'}$ WZW model, the former corresponds to the periodicity $k-k' \ (\text{mod}.N)$ and the latter produces the periodicity $k+k'\ (\text{mod}.N)$. The theory $M_{\text{UV}}\otimes \overline{M_{\text{IR}}}$ or $M_{\text{UV}}\otimes M_{\text{IR}}$ has nonchiral or chiral (half) integer spin simple current corresponding to anomaly matching or anomaly cancellations. 
For anomaly matching flow, we propose the following coset decomposition or Witt equivalence,
\begin{equation}
M_{\text{UV}}=\frac{M_{\text{UV}}}{ M_{\text{IR}}} \otimes_{\text{D}} M_{\text{IR}}
\end{equation}
or equivalently,
\begin{equation}
M_{\text{IR}}=\frac{\overline{M_{\text{UV}}}}{ \overline{M_{\text{IR}}}} \otimes_{\text{D}} M_{\text{UV}}
\end{equation}
where the product $\otimes_{\text{D}}$ is a kind of Deligne product.

For the anomaly cancellation flow, we also propose the following Witt equivalent relation, 
\begin{equation}
M_{\text{UV}}=(  M_{\text{UV}}\otimes_{\text{D}} M_{\text{IR}}) \otimes_{\text{D}} \overline{M_{\text{IR}}}
\end{equation}
or equivalently,
\begin{equation}
\overline{M_{\text{IR}}}=(  \overline{M_{\text{IR}}}\otimes_{\text{D}} \overline{M_{\text{UV}}}) \otimes_{\text{D}} M_{\text{UV}}
\end{equation}
where the product is a Deligne product and the object $M_{\text{UV}}\otimes_{\text{D}} M_{\text{IR}}$ appears in recent literature \cite{Kikuchi:2024cjd,Fukusumi:2024ejk}. The corresponding decomposition has been studied as the level-rank duality\cite{Kuniba:1990im,Kuniba:1990zh,Nakanishi:1990hj} and the method has applied to the $SU(N)$ fractional Hall states in \cite{Bourgine:2024ycr}. 

\begin{figure}[htbp]
\begin{center}
\includegraphics[width=0.5\textwidth]{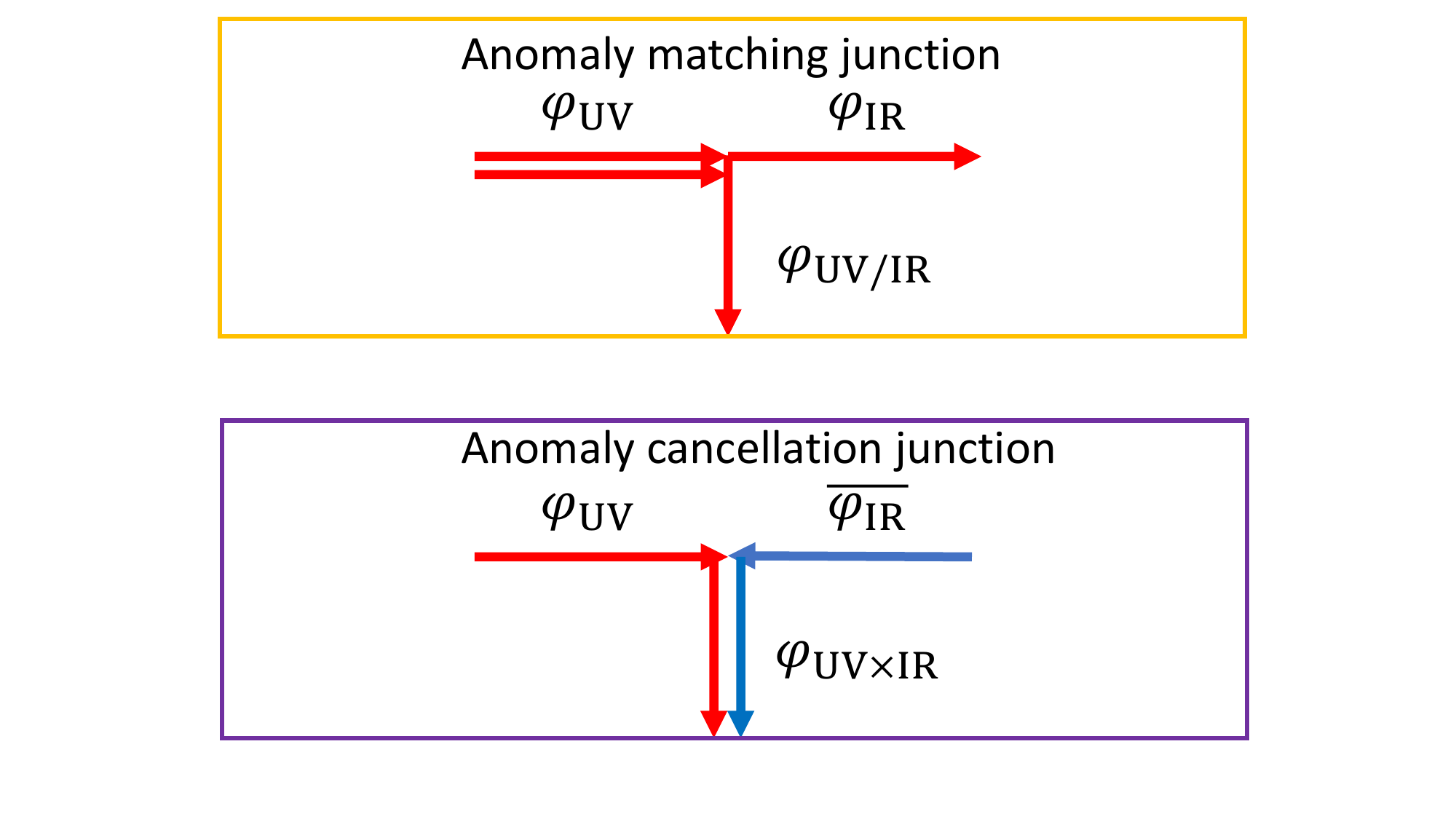}
\caption{Domain wall or tri-wire junction problem corresponding to the massless RG. The fields $\varphi$ and $\overline{\varphi}$ are chiral or antichiral fields, respectively. An exact description of the theory appearing in the domain wall is an open problem in general, but there exists respective research verifying this picture (anomaly matching case, in particular)\cite{Quella:2006de,Kimura:2014hva,Kimura:2015nka}.}
\label{anomaly_junction}
\end{center}
\end{figure}

By considering the corresponding quantum Hall point contact or quantum tri-wire junction problem\cite{Oshikawa:2005fh}, it is possible to interpret the decomposition more phenomenologically(Fig. \ref{anomaly_junction}). The anomaly matching flow corresponds to the situation where the current flows from one wire to another wire with some dissipation at the junction. The anomaly cancellation flow corresponds to the case where the currents from one wire and another wire cancel at the junction point. In the main text, we have paid particular attention to the symmetry-preserving domain wall, but the coset representation will contain sectors corresponding to the charged domain wall. If there exists a correspondence between charged sectors and uncharged sectors both in UV and IR (or both systems have fractional supersymmetry), one can straightforwardly construct a mapping relation between UV and IR theories by shifting the charge appearing in the symmetry-preserving domain wall\cite{Fukusumi:2024ejk}.

For the $Z_{N}$ symmetric model, the (half) integer spin simple current should govern the classification of $Z_{N}$ symmetry-preserving domain walls. As a straightforward generalization, nonsimple current\cite{Fuchs:2000gv} with integer spin, which can be called integer spin nonsimple current, should govern the anomaly classification of noninvertible symmetric models, and this is related to maverick coset in the literature\cite{Dunbar:1993hr,Pedrini:1999iy,Frohlich:2003hm,Cordova:2023jip}. Because of its noninvertible nature, it seems necessary to study the Deligne product structure or chiral-chiral chiral-antichiral anyon condensations of such theories more carefully. Related to this direction, it is also known that, if two systems have the same fusion rule, particular types of modular invariants appear\cite{Gannon:1992np,Gannon:1998rw}.

By extending the notion of anomaly classification to noninvertible symmetry, one can expect the appearance of a noninvertible symmetry-protected topological phase. We list several recent literature in this research direction\cite{Seifnashri:2024dsd,Jia:2024zdp,Lu:2025gpt,Cao:2025qhg,Seifnashri:2025fgd,Aksoy:2025rmg,Lu:2025rwd,Furukawa:2025flp}. The corresponding structure should also be governed by the integer spin nonsimple current. We expect that further anomaly classification of noninvertible symmetry and the analysis of the corresponding massless RGs are important for this research direction \cite{Kaidi:2023maf,Nakayama:2024msv}.

For an intuitive understanding, we demonstrate the naturality of the noninteger coefficient by studying implications of the quantum dimensions of the anyons at the domain wall (for readers interested in more general settings, see the recent work \cite{Fukusumi:2025xrj} by the first author). First, we assume the decomposition of the UV theory to $\mathbf{A}_{(1)}=\mathbf{A}_{\text{DW}}\otimes \mathbf{A}_{(2)}$ where $\mathbf{A}_{\text{DW}}$ represents the theory at the domain wall. These arguments are applicable to both the anomaly cancellation and anomaly matching cases, and we drop the information on the chirality of the objects for simplicity. In this setting, one can obtain the following tensor decomposition of the anyonic objects,
\begin{align}
\mathbf{A}_{(1)}&=\mathbf{A}_{\text{DW}}\otimes \mathbf{A}_{(2)}, \\
a_{(1)}&=\sum_{a_{\text{DW}}, a_{(2)}}B^{\alpha_{(1)}}_{a_{\text{DW}}, a_{(2)}}a_{\text{DW}}\otimes a_{(2)}
\end{align}
where $a$ is the anyons in the corresponding fusion ring and $B$ is the matrix specifying the tensor decomposition. Here, we also assume the quantum dimension $q_{a_{\text{DW}}}$ on the domain wall theory $\mathbf{A}_{\text{DW}}$ is well-defined. The quantum dimension can be understood as ring homomorphism $d: \mathbf{A}_{\text{DW}}\rightarrow \mathbb{C}$\cite{Fuchs:1991ci,Fuchs:1993et}. Hence, the algebraic relation is compatible with the replacement of the symbol $a_{\text{DW}}$ by $q_{a_{\text{DW}}}$. One can obtain the ring homomorphism as,
\begin{align}
\rho=I\otimes d: \mathbf{A}_{(1)}&\rightarrow \mathbf{A}_{(2)}, \\
\rho(a_{(1)})&=\sum_{a_{\text{DW}}, a_{(2)}}B^{\alpha_{(1)}}_{a_{\text{DW}}, a_{(2)}} q_{a_{\text{DW}}} a_{(2)}.
\end{align}
By introducing the new matrix $A^{a_{(1)}}_{a_{(2)}}=\sum_{a_{\text{DW}}}B^{\alpha_{(1)}}_{a_{\text{DW}}, a_{(2)}} q_{a_{\text{DW}}}$, one can obtain the form 
\begin{equation}
\rho(a_{(1)})=\sum_{a_{(2)}}A^{a_{(1)}}_{a_{(2)}}a_{(2)}.
\end{equation}
It should be stressed that even when assuming the NIM-rep of $B$, the resultant coefficient matrix $A$ can be outside of NIM-rep, because the quantum dimension can take noninteger values. Hence, the existence of the coset representation, or level-rank duality\cite{Kuniba:1990im,Kuniba:1990zh,Nakanishi:1990hj}, inevitably induces the fusion ring homomorphisms with noninteger coefficients in general. The hierarchical strucutre of related to the level-rank duality has been studied in \cite{Bourgine:2024ycr}, but this research direction is relatively less familiar.

\section{Noninteger coefficient in the sandwich construction}
In this section, we revisit the inevitable appearance of the noninteger coefficient in studying fermionic chiral CFT or superfusion category theory, when studying their relationship to the corresponding bosonic models. For simplicity, we concentrate on the Majorana chiral CFT, and we note an established review\cite{Ginsparg:1988ui}. We note that the corresponding category theory, premodular fusion category, is still under development, but its fundamental algebraic data, fusion ring, has already been obtained in a way consistent with the sandwich construction\cite{Fukusumi:2024cnl,Fukusumi:2025ljx}.

The chiral Majorana CFT has the four objects $\{ I',\psi', e',m'\}$, satisfying the following relations known as the double semion fusion rule,
\begin{align}
\psi'\times \psi'&=I' \\
e'\times e'&=m' \times m'=I' \\
\psi'\times e'&=m' 
\end{align}
Recently, this fusion rule has been revisited in \cite{Huang:2023pyk,Wen:2024udn,Fukusumi:2024cnl,Bhardwaj:2024ydc,Huang:2024ror} by studying their reduction from the bulk CFT, $\{I, \psi, \overline{\psi}, \epsilon, \sigma_{\text{Bulk},+}, \mu_{\text{Bulk},+}\}$, which can be identified as the $Z_{2}$ extension of the Ising spherical fusion category $\{ I, \epsilon,\sigma_{\text{Bulk},+}\}$.

In some of the recent references, for example, in \cite{Huang:2023pyk,Bhardwaj:2024ydc,Huang:2024ror}, categorical derivations of the reduction from the bulk fusion rule (spherical fusion category), the chiral fusion rule havea been discussed. The following is the main reduction in the recent literature,
\begin{align} 
I' &\sim I+\epsilon, \\
\psi' &\sim \psi +\overline{\psi}, \\
e' &\sim \sigma_{\text{Bulk},+}, \\
m' &\sim \sigma_{\text{Bulk},+}.
\end{align}
The categorical arguments provide an intuitive understanding of the phenomena, but the above expression is misleading for a wider audience or from a perspective emphasizing the established mathematics. We also note the above symbol $\sim$ is sometimes replaced by $=$, without explicit algebraic demonstrations.

First, the righthandside should be defined by the $Z_{2}$ extended SFC\cite{etingof2009weaklygrouptheoreticalsolvablefusion}, but the disordered field $\mu_{\text{Bulk}}$ which \emph{should} be present in a consistent fermionic theory is absent. Hence, the counting of the Ramond sector is incorrect. Moreover, the identification $e \sim \sigma_{\text{Bulk}}, m \sim \sigma_{\text{Bulk}}$ are not compatible with the double semion fusion ring with $e\neq m$ (We note the abuse of the ideas in anyon condensation\cite{Bais:2008ni} does not resolve this issue). The counting of the Ramond sector is  called fermionic zero modes, and this structure has played siginificant roles in the studies of supersymmetric models. However, we also note that some of the established literature often contains misleading arguments overlooking the corresponding zero modes $\mu_{\text{Bulk},+}$, and this problem in the recent literature is understandable from this point. 

Second, the fusion product of the right-hand side is not compatible with the fusion product of the left-hand side. For example, one can obtain the relation $(I+\epsilon)\times (I+\epsilon)=2(I+\epsilon)$, but this is not compatible with the standard relation $I'\times I' =I'$ or the definition of the unity. One might say that the coefficient $2$ is negligible in the representation theory, but this is not true without careful arguments. Moreover, this number $2$ can be identified as edge modes (or boundary qubit) of symmetry protected topological orders\cite{Graham:2003nc,Fukusumi:2020irh,Okada:2024qmk}.

Third, because of the first and second problems, it is difficult to obtain the fusion coefficient of general chiral CFT from the existing data of the bulk CFT. Except for the works by the first author and collaborators\cite{Fukusumi:2024cnl,Fukusumi:2025ljx}, the systematic construction of the corresponding fusion coefficients in a general setting has not been studied. We note that there exist a research direction on their formal categorical structure\cite{Etingof:2009yvg}, but the corresponding algebraic reduction is difficult to read.

One can implement the algebraic resolution of the above problems straightforwardly by introducing the following identifications,
\begin{align} 
I' &= \frac{I+\epsilon}{2}, \\
\psi' &= \frac{\psi +\overline{\psi}}{2}, \\
e' &= \frac{\sigma_{\text{Bulk},+}}{\sqrt{2}}, \\
m' &= \frac{\mu_{\text{Bulk},+}}{\sqrt{2}}.
\end{align}
The right handsides perfectly reproduce the algebraic relation of the left hand sides, by identifying $\{ I', \psi', e', m'\}$ as a subalgebra of $\{I, \psi, \overline{\psi}, \epsilon, \sigma_{\text{Bulk},+}, \mu_{\text{Bulk},+}\}$. This interpretation of the reduction by taking a subalgebra is also compatible with the intimate relationship between chiral CFT, BCFT, and bulk renormalization group in the literature. 

We also note that the chiral Ising fusion rule, $\{ I', \psi', \sigma' \}$ can be reproduced from the chiral Majorana fusion rule by the identification $\sigma'=\frac{e'+m'}{\sqrt{2}}$. One can easily check the resultant nonabelian fusion rule, $\sigma' \times \sigma'=I'+\psi'$. This kind of identification appeared commonly in literature on supersymmetric models, and the resultant category theory is sometimes formulated as $\mathbb{C}$-linear category or superfusion category.

\section{Verlinde operator and topological defects: $\mathbb{C}$-linear category and fusion category}
In this section, for readers unfamiliar with quantum Hamiltonian formalism of QFTs, we briefly review the arguments in the Zuber-Petkova\cite{Petkova:2000ip} from a modern perspective. First, we concentrate our attention on the $A$-type diagonal minimal model given by the following partition functions:
\begin{equation}
Z(\tau,\overline{\tau})=\sum_{a}\chi_{a}(\tau)\overline{\chi_{a}}(\overline{\tau})
\end{equation}
where $a$ is the label of the primary fields, and $\tau$ and $\overline{\tau}$ are holomorphic or antiholomorphic modular parameter and $\chi$ is the corresponding characters.

Corresponding to this character, one can introduce the following projection operators,
\begin{equation}
P_{a}=\sum_{M,\overline{M}} |a,M\rangle \langle a,M| \otimes \overline{|a,\overline{M}\rangle} \overline{\langle a,\overline{M}|}
\end{equation}
where $M$ and $\overline{M}$ are the labels of the chiral or antichiral descendant fields, corresponding to the Verma module. The most fundamental point in this representation is, by definition, $P_{a}$ is commutative with the CFT Hamiltonian $H_{CFT}$, i.e. $[H_{CFT},P_{a}]=0$. In other words, the above form provides representations of conserved charges. Hence, the operators form a  ring over $\mathbb{C}$. Moreover, by changing the basis by a linear transformation, one can introduce the following topological symmetry operators,
\begin{equation}
Q_{a}=\sum_{b}\frac{S_{a,b}}{S_{I,b}} P_{b}
\end{equation}
where $S$ is the so-called modular $S$ matrix defined by $\chi_{a}(\tau)=\sum_{b}S_{a,b}\chi_{b}(-1/\tau)$. By applying the Verlinde formula\cite{Verlinde:1988sn}, one can check that the topological symmetry operator satisfies the relation,
\begin{equation}
Q_{a}\times Q_{b}=\sum_{c}N_{ab}^{c}Q_{c}
\end{equation}
by identifying $\times$ as the multiplication of the linear operators. By applying the modular $S$ transformation, the operators produce defects satisfying the fusion rules. Hence, the topological symmetry operator provides a natural basis to relate defects described by a fusion category and conserved charges described by a ring over $\mathbb{C}$. However, this does not imply that their definitions are compatible with RG flows straightforwardly, as we have demonstrated in the main text. Because RG flow is nothing but the addition of the operator to the quantum Hamiltonian, and this addition changes the role of space and time. Hence, the roles of the defects and conserved charges are deformed nontrivially through the process in general.

\clearpage

\bibliographystyle{ytphys}
\bibliography{homomorphism_Gaiotto}

\end{document}